\documentclass[10pt,preprint]{aastex}

\usepackage{amsmath,natbib,graphicx}

\def\Cpp{{\textit{C{\small++}}}\ }
\def\Cosmos{{\textit{Cosmos}}\ }
\def\Cosmospp{{\textit{Cosmos{\small++}}}\ }
\def\Cosmosppbf{{\textbf{\textit{Cosmos{\small++}}}}\ }

\begin{document}

\title{ \Cosmosppbf: Relativistic Magnetohydrodynamics
                     on Unstructured Grids with Local Adaptive Refinement }

\author{Peter Anninos\altaffilmark{1},
P. Chris Fragile\altaffilmark{2}, and
Jay D. Salmonson\altaffilmark{1}}
\altaffiltext{1}{University of California,Lawrence Livermore National
Laboratory, Livermore CA 94550 }
\altaffiltext{2}{Department of Physics, University of California, Santa
Barbara, CA 93106; fragile@physics.ucsb.edu}

\date{{\small    \today}}
\date{{\small   \LaTeX-ed \today}}

\begin{abstract}
A new code and methodology are introduced for solving the general relativistic
magnetohydrodynamic (GRMHD) equations in fixed background
spacetimes using time-explicit, finite-volume
discretization.  The code has options for solving
the GRMHD equations using traditional artificial-viscosity (AV) or
non-oscillatory central difference (NOCD) methods,
or a new extended AV (eAV) scheme using
artificial-viscosity together with a dual energy-flux-conserving formulation.
The dual energy approach allows for accurate modeling of highly relativistic
flows at boost factors well beyond what has been achieved to date
by standard artificial viscosity methods. It provides the benefit of Godunov
methods in capturing high Lorentz boosted flows but without complicated Riemann
solvers, and the advantages of traditional artificial viscosity methods in their
speed and flexibility. Additionally, the GRMHD equations are solved on an
unstructured grid that supports local adaptive mesh refinement using a
fully threaded oct-tree (in three dimensions) network
to traverse the grid hierarchy across levels and immediate neighbors.
A number of tests are presented to demonstrate robustness of the
numerical algorithms and adaptive mesh framework over a wide spectrum of
problems, boosts, and astrophysical applications,
including relativistic shock tubes, shock collisions,
magnetosonic shocks, Alfv\'en wave propagation, blast waves,
magnetized Bondi flow, and the magneto-rotational instability in
Kerr black hole spacetimes.
\end{abstract}


\section{Introduction}
\label{sec:intro}

Magnetohydrodynamic (MHD) driven processes in high-energy astrophysical
environments can be extremely difficult to model numerically, especially
where flows are strongly trans-sonic, characteristic speeds approach the
speed of light, or nonlinear effects couple over short temporal
and large spatial dynamical ranges. For example, studies of phase transitions
in the early universe \citep{fra03a}, and the formation of primordial magnetic
fields and black holes demand resolution across many decades of scale
to model phase interactions, wall dynamics, and microphysical coupling
effects. Similarly, simulations of black hole accretion flows
\citep[e.g. ][]{fra05b}, thought to exist in quasars, active galactic
nuclei, X-ray binaries, core-collapse supernovae, and gamma-ray bursts,
require stably evolving the GRMHD equations in a strongly curved
background spacetime over many dynamical scales.
Addressing such problems requires numerical algorithms that are robust enough
to simulate ultra-relativistic waves and shocks, and accurate enough to treat
various physical instabilities that can be seeded at very small amplitudes and
frequencies and yet affect global solutions.  We also desire algorithms that can
capture the propagation of outflows or highly
compressed features beyond the limits imposed by single mesh simulations.

A number of codes have been developed in recent years to solve both the
special and general relativistic MHD equations. Our approach is most similar to
that taken by \cite{dev03a}, which is based on an internal energy formulation
with artificial viscosity (AV) for shock capturing and a method
of characteristics for the magnetic fields. However, other approaches include
the simplified total variation diminishing (sTVD) method \citep{koi99}, and
approximate Riemann solver schemes \citep{kom99,gam03a,duez05,anton05}
based on fully conservative formulations of the MHD equations. Riemann methods
are significantly more complex than AV schemes, but are considered more robust
for ultra-relativistic problems. Although viscosity-based methods are generally
accurate, fast, simple to implement, and easy to expand to include multi-physics
capabilities, they have historically been limited in their range of applicability
to moderately relativistic flows, with boost factors less than a few.

In this paper we describe a new massively parallel, multi-dimensional numerical
code \Cosmospp with improved (over traditional AV methods) shock capturing
capabilities in the high boost regime and adaptive mesh refinement (AMR), representing
a significant advance over our previous numerical code
\citep[\Cosmos,][]{ann03a}.
\Cosmospp introduces a more complex unstructured mesh system that allows for
arbitrarily connected hexahedral (quadrilateral in 2D) cells to conform to any
boundary or shape. It also allows for dentritic-type zoning characteristic
of refined grids at level interfaces. Our AMR framework differs
from the more standard approach \citep{ber84,ber89} by refining individual cells
rather than introducing patches of sub-grids composed of multi-dimensional
arrays of cells, thus providing greater flexibility in modeling complex
flows and greater efficiency in positioning computational resources.
This framework is similar to that described by \cite{kho98}, but we generalize
the method to unstructured meshes. Also, as in our predecessor code, several
schemes have been implemented to solve the hydrodynamical equations,
including both artificial viscosity and non-oscillatory
central difference (NOCD) methods. However, here we use discrete finite-volume
methods in place of finite differences due to the unstructured nature of
the grid. We also introduce in this paper a new dual energy procedure (eAV)
that allows conventional artificial viscosity methods
to be extended and work robustly at arbitrarily high Lorentz factors.

The basic equations, numerical methods, and tests of the code are
described in the remaining sections. Although \Cosmospp supports many
different physics packages in both covariant Newtonian and general
relativistic systems, including chemical networks, flux-limited radiation
diffusion, self-gravity, magnetic fields, and radiative cooling, only the
GRMHD algorithms in the AMR framework are covered in this paper.
The Newtonian physics capabilities are currently the same
as those of \Cosmos, which were presented in \citet{ann03b}
and \citet{fra05a}.

\section{Basic Equations}
\label{sec:equations}

Two separate formulations of the GRMHD equations are presented in this section.
The first form incorporates an evolution equation for the internal energy
that is appropriate for schemes using artificial viscosity methods for
capturing shocks \citep{wil72,wil79,haw84b}. The second form is derived
directly from the primitive
form of stress-energy conservation and is thus fully conservative
and provides the basis for the non-oscillatory central difference schemes.
We use the standard notation in which 4(3)-dimensional tensor quantities are
represented by Greek(Latin) indices.

For a perfect fluid, the stress-energy tensor is generated from a linear
sum of the hydrodynamic $T^{\mu\nu}_{H}$ and magnetic contributions $T^{\mu\nu}_{B}$
\begin{eqnarray}
T^{\mu\nu} &=& T^{\mu\nu}_{H} + T^{\mu\nu}_{B}  \nonumber \\
           &=& \rho h u^\mu u^\nu + P g^{\mu\nu}
             + \frac{1}{4\pi}\left(\frac{1}{2} g^{\mu\nu} ||B||^2 +
                                   u^\mu u^\nu ||B||^2 - B^\mu B^\nu\right)
             \label{eqn:tmn} \\
           &=& \left(\rho h + 2P_B\right) u^\mu u^\nu
               + \left(P + P_B\right) g^{\mu\nu} - \frac{1}{4\pi} B^\mu B^\nu
               \nonumber
           ~,
\end{eqnarray}
where
\begin{equation}
h = 1 + \epsilon + \frac{P}{\rho} + \frac{||Q||}{\rho}
  = 1 + \Gamma\epsilon + \frac{||Q||}{\rho}
\label{eqn:enthalpy}
\end{equation}
is the relativistic enthalpy,
$\epsilon$ is the specific internal energy, $P$ is the fluid pressure,
$||Q||$ is the bulk scalar (artificial) viscosity,
$P_B = ||B||^2/8\pi = g_{\mu\nu} B^\mu B^\nu/8\pi$ is the
magnetic pressure, and $B^\mu$ is the magnetic induction in the rest frame of the fluid.
Also, $u^\mu$ is the contravariant 4-velocity, $g_{\mu\nu}$ is the 4-metric, and
$\Gamma$ is the adiabatic index assuming an ideal gas equation of
state $P = (\Gamma-1)\rho\epsilon$.

\subsection{Internal Energy Formulation}
\label{sec:internal_e}

This method uses a form of the GRMHD equations similar
to \citet{dev03a}, derived from velocity normalization $u_\mu u^\mu = -1$,
baryon conservation $\nabla_\mu(\rho u^\mu) = 0$, energy conservation
$u_\nu \nabla_\mu T^{\mu\nu} = 0$, momentum conservation
$(g_{\alpha\nu} + u_{\alpha} u_{\nu}) \nabla_\mu T^{\mu\nu} = 0$,
and magnetic induction $\nabla_\mu(u^\mu B^\nu - B^\mu u^\nu) = 0$.
Nevertheless, it is worth explicitly writing these
equations out for clarity, since we employ a unique expansion
and grouping of the terms in our numerical implementation.
In flux-conserving form, the evolution
equations for mass, internal energy, momentum, and magnetic induction are:
\begin{eqnarray}
 \partial_t D + \partial_i (DV^i) &=& 0 ~,  \label{eqn:av_de} \\
 \partial_t E + \partial_i (EV^i) &=&
    - \left(P + k_{\dot{W}} ||Q|| \right) \partial_t W
    - \left(P \delta^j_i + Q^j_i\right) \partial_j (WV^i) ~,
    \label{eqn:av_en} \\
 \partial_t S_j + \partial_i (S_j V^i) &=&
      \frac{1}{4\pi} \partial_t (\sqrt{-g} B_j B^0)
    + \frac{1}{4\pi} \partial_i (\sqrt{-g} B_j B^i)
    \label{eqn:av_mom} \nonumber \\
    & & {} + \left( \frac{S^\mu S^\nu}{2S^0} - \frac{\sqrt{-g}}{8\pi}
             B^\mu B^\nu \right) \partial_j g_{\mu\nu}
    - \sqrt{-g}~\partial_i ( (P + P_B)\delta^i_j + Q^i_j ) ~, \\
 \partial_t \mathcal{B}^j + \partial_i (\mathcal{B}^j V^i) &=&
    \mathcal{B}^i \partial_i V^j + \eta~\partial^j (\partial_i \mathcal{B}^i) ~,
      \label{eqn:av_ind}
\end{eqnarray}
where $g$ is the 4-metric determinant, $\delta^i_j$ is the Kronecker delta tensor,
$W=\sqrt{-g} u^0$ is the relativistic boost factor, $D=W\rho$ is the generalized fluid density,
$V^i=u^i/u^0$ is the transport velocity,
$S_\mu = W(\rho h + 2P_B) u_\mu$ is the covariant momentum density,
$E=We=W\rho\epsilon$ is the generalized internal energy density,
$Q^i_j$ is the tensor artificial viscosity used for shock capturing,
$k_{\dot{W}}$ is a switch used to activate a viscosity multiplier
for the $\partial_t W$ source,
and $\eta$ is a coefficient related to the largest characteristic speed in the
flow multiplying the divergence cleanser function.
Notice there are two representations of the magnetic field in these equations:
$B^\mu$ is the rest frame magnetic field 4-vector defined in the stress tensor
definition (\ref{eqn:tmn}) and
\begin{equation}
\mathcal{B}^\mu = W(B^\mu - B^0 V^\mu)
\end{equation}
is the divergence-free ($\partial \mathcal{B}^i / \partial x^i = 0$),
spatial ($\mathcal{B}^0=0$) representation of the field.  The time
component of the magnetic field $B^0$ is recovered from the orthogonality
condition $B^\mu u_\mu = 0$
\begin{eqnarray}
B^0 & = & -\frac{g_{0i} B^i + g_{ij} B^j V^i}{g_{00} + g_{0i} V^i}
        = -\frac{g_{0i} \mathcal{B}^i + g_{ij} \mathcal{B}^j V^i}
              {W(g_{00} + 2 g_{0i} V^i + g_{ij} V^i V^j)} \\
    & = & \frac{\mathcal{B}^i S_i}{\sqrt{-g}W(\rho h + 2P_B)}
~,
\label{eqn:B0}
\end{eqnarray}
and we use
\begin{equation}
P_B = \frac{1}{8\pi} g_{\mu\nu} B^\mu B^\nu
    = \frac{1}{8\pi} \frac{g_{ij} {\cal B}^i {\cal B}^j}{W^2} +
      \frac{1}{8\pi}\left(\frac{B^0 \sqrt{-g}}{W}\right)^2
\end{equation}
to compute the magnetic pressure from the divergence-free field.

We note that the term $\partial_i(\sqrt{-g} B_j B^i)$ in equation (\ref{eqn:av_mom})
can be cast into projected and divergence components
$B^i\partial_i(\sqrt{-g} B_j) + \sqrt{-g} B_j \partial_i B^i$.
The advantage of this form is that the method of characteristics
described by \cite{sto92} can be applied easily to the projected component
after generalizing the definitions of density and momentum accounting
for relativistic inertia and boost contributions.
This procedure is important for achieving stable simulations of sheared Alfv\'en waves,
as demonstrated in \S\ref{subsubsec:alfven}.

Two additional sets of equations are needed for the transport velocity $V^i$ and
boost factor $W$. The velocity is derived from the 4-momentum normalization
$S_\mu S^\mu = W^2(\rho h +2P_B)^2 u_\mu u^\mu = -W^2(\rho h + 2P_B)^2$, from which we
compute
\begin{equation}
S_0 = -\frac{g^{0i} S_i}{g^{00}}
      +\frac{1}{g^{00}} \left[ \left(g^{0i} S_i\right)^2 -
                               g^{00} \left(W^2\left(\rho h + 2 P_B\right)^2 +
                                            g^{ij} S_i S_j\right)
                        \right]^{1/2}
~,
\end{equation}
and  then $V^i = S^i/S^0$. A convenient formula is derived for the
boost factor using $u^0$ obtained from the 4-velocity normalization written as
$u_\mu u^\mu = u^0 V^\mu S_\mu/(W(\rho h+2P_B)) = -1$. The boost ($W=\sqrt{-g} u^0$)
is then evaluated in one of two ways
\begin{equation}
W = -\frac{\sqrt{-g}~(-S_\mu S^\mu)^{1/2}}{S_\mu V^\mu}
  = \frac{\sqrt{-g}~S^0}{(-S_\mu S^\mu)^{1/2}} ~,
\end{equation}
where $V^0=1$, $S^0$ is computed from the 4-momentum normalization described above,
and the other fields are known from solutions to the evolution equations.

\subsection{Conservative Energy Formulation}
\label{sec:conservative_e}

A second class of numerical methods presented in this paper are based
on a conservative hyperbolic formulation of the GRMHD equations.
It is the same approach used by \citet{ann03a}, except here we include
magnetic fields. The equations are derived directly from the conservation of
stress-energy $\nabla_\mu T^{\mu\nu} = 0$ and then
decomposed into space and time components
\begin{equation}
\partial_t (\sqrt{-g}~T^{0\nu}) + \partial_i (\sqrt{-g}~T^{i\nu}) = \Sigma^\nu ,
\label{eqn:tmnu3}
\end{equation}
with curvature source terms
\begin{equation}
\Sigma^\nu = -\sqrt{-g}~T^{\beta\gamma}~\Gamma^\nu_{\beta\gamma}.
\end{equation}
The form of the differential equations that we solve, derived after
substituting the perfect fluid stress tensor (\ref{eqn:tmn}) into
(\ref{eqn:tmnu3}), are
\begin{eqnarray}
 \partial_t {\cal E} + \partial_i ({\cal E}V^i)
                     + \partial_i F^{0i} &=& \Sigma^0 ~,
                     \label{eqn:hr_en} \\
 \partial_t {\cal S}^j + \partial_i ({\cal S}^j V^i)
                     + \partial_i F^{ij} &=& \Sigma^j ~,
                     \label{eqn:hr_mom}
\end{eqnarray}
where we have explicitly split off the transport term from the other
divergence flux contributions $F^{i\alpha}$, defined as
\begin{equation}
 F^{i\alpha} = \sqrt{-g}~\left( (g^{i\alpha} - g^{0\alpha} V^i)~(P+P_B)
                               - \frac{1}{4\pi}(B^i B^\alpha - B^0 B^\alpha V^i)
                         \right) ~.
\end{equation}
The variables $D$, $V^i$, and $g$ are defined as in
section \S\ref{sec:internal_e}, and
\begin{eqnarray}
{\cal E}   = \sqrt{-g} T^{00} &=&  \frac{W^2}{\sqrt{-g}} (\rho h + 2P_B)
                                 + \sqrt{-g}~g^{00} (P+P_B)
                                 - \frac{1}{4\pi} \sqrt{-g} B^0 B^0 ~, \\
{\cal S}^j = \sqrt{-g} T^{0j} &=&  \frac{W^2}{\sqrt{-g}} (\rho h + 2P_B) V^j
                                 + \sqrt{-g}~g^{0j} (P+P_B)
                                 - \frac{1}{4\pi} \sqrt{-g} B^0 B^j ~, \\
\end{eqnarray}
are the new expressions for energy and momentum. We also note that the
divergence-free magnetic induction equation (\ref{eqn:av_ind}) can be written
in the fully conservative form
\begin{equation}
\partial_t \mathcal{B}^j + \partial_i (\mathcal{B}^j V^i - \mathcal{B}^i V^j)
                         = \eta~\partial^j (\partial_i \mathcal{B}^i) ~,
\label{eqn:hr_ind}
\end{equation}
as required by the central difference schemes described in \S\ref{subsec:nocd}.

It is convenient to express ${\cal E}$ and ${\cal S}^i$ in terms
of the internal energy fields $E$ and $S_\alpha$
\begin{eqnarray}
{\cal E}   &=&
             \frac{W^2}{\sqrt{-g}}\left(\frac{D}{W} + \Gamma\frac{E}{W} + 2P_B\right)
             + \sqrt{-g}~g^{00}\left( (\Gamma-1) \frac{E}{W} + P_B\right)
             - \frac{1}{4\pi} \sqrt{-g} B^0 B^0 ~, \\
{\cal S}^j &=&
             g^{j\alpha} S_\alpha
             + \sqrt{-g}~g^{0j}\left( (\Gamma-1) \frac{E}{W} + P_B\right)
             - \frac{1}{4\pi} \sqrt{-g} B^0 B^j ~.
\end{eqnarray}
The inverse energy relation
\begin{eqnarray}
\frac{E}{W} = \left( \frac{{\cal E} \sqrt{-g}}{W^2} - \frac{D}{W} - 2P_B
                     - \left(\frac{\sqrt{-g}}{W}\right)^2
                       \left(g^{00} P_B - \frac{B^0 B^0}{4\pi} \right)
              \right)
              \frac{1}{\Gamma + (\Gamma-1) g^{00} (\sqrt{-g}/W)^2}
\label{eqn:hr_pressure}
\end{eqnarray}
is useful for computing the fluid pressure ($P = (\Gamma-1)E/W$), and for
comparing relative errors in the dual energy update procedure described
in \S\ref{sec:methods}.

\section{Numerical Methods}
\label{sec:methods}

\begin{figure}
\plotone{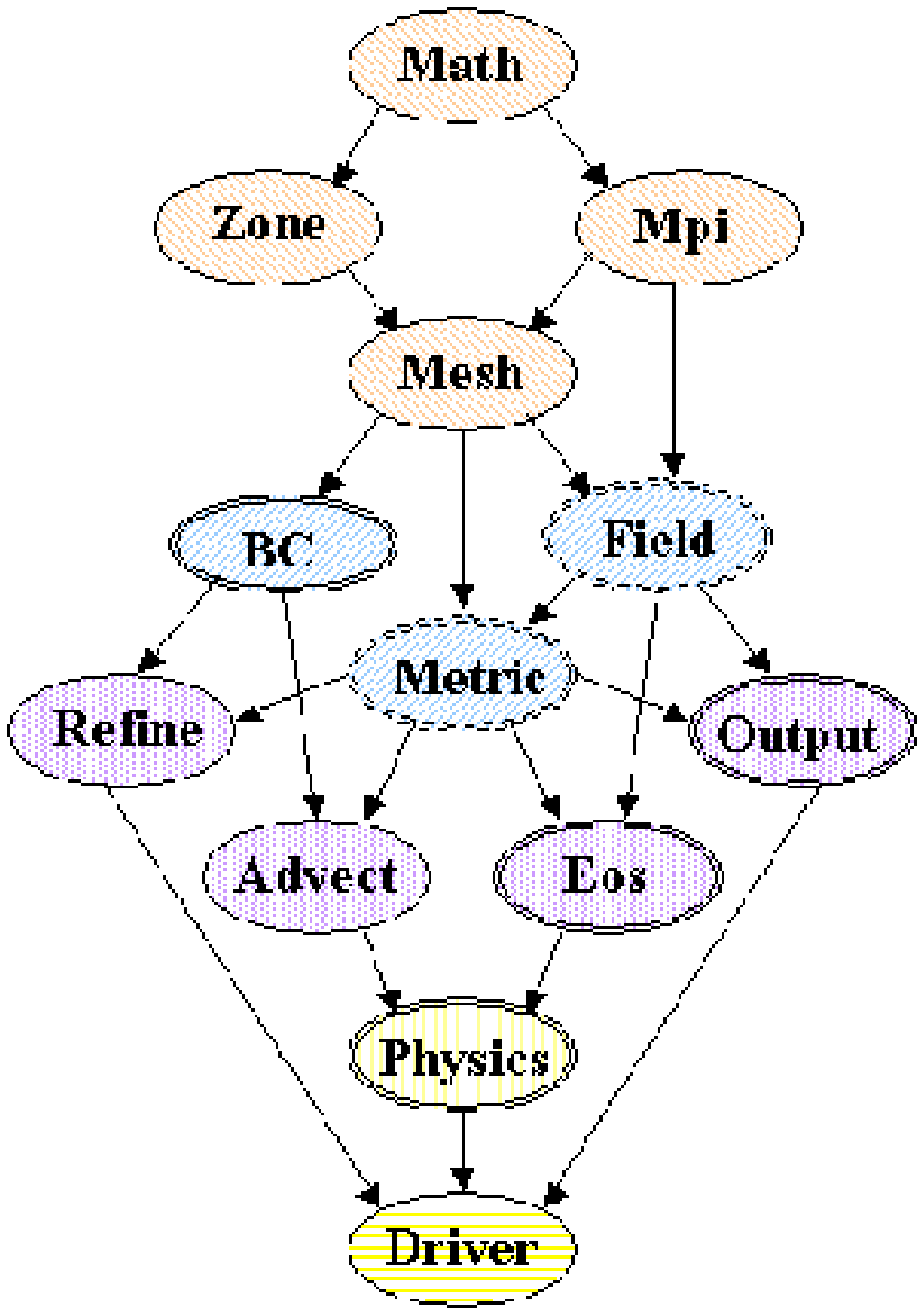}
\caption{Flowchart describing the relative dependencies of the
basic classes in our code design. The core, or lowest level,
functions include the abstract mathematical (e.g., vector, tensor, etc...)
constructs, MPI domain exchange functions, zone attributes
defining local cell geometries and connectivity, and the
global mesh construction. At the intermediate
level we have the polymorphic boundary condition objects, field
operators (e.g., gradients), and the metric functions defining the
proper local cell geometry when using curvilinear coordinates.
The highest level classes are the
different physics packages which also depend on additional
input from microphysics (e.g., equations of state, opacities)
and advection treatments. The physics objects, refinement functions,
and output options are incorporated into a main driver class that
controls the evolution, mesh refinement, and dump sequence, and
generally allows the user to interactively ``steer'' the simulations.
}
\label{fig:newfig1}
\end{figure}

\Cosmospp is designed using object-oriented principles to create
mathematical abstraction classes for vector and tensor (both
three and four dimensional) objects on which most functional operations
are based. \Cosmospp also takes advantage of the operator overload, inheritance,
polymorphism and virtual methods features of the \Cpp language in its
design of all the basic class structures. This simplifies the user interface
considerably, reduces the amount of coding, and allows for the code to be
easily developed and expanded.
Figure \ref{fig:newfig1} illustrates the relative dependencies of the
basic classes, ranging from the fundamental math, MPI, zone and mesh classes,
to the intermediate structures defining the metric geometry, field
operators (e.g., gradients, filters, etc...), and boundary conditions,
leading finally to the higher level classes where the physics, output,
refinement, and evolution driver algorithms depend on all the
lower level functions. The classes outlined with solid double lines
in Figure \ref{fig:newfig1} represent base classes where virtual methods
functions proved advantageous in their polymorphic design. Classes outlined
with double dashed lines indicate where polymorphism can help
simplify and improve the code design as we add new options
for background metrics and gradient operators.

All data structures (mesh and fields) are
stored as vector or map container classes using the \Cpp
Standard Template Library (STL) which provides convenient
memory access functions and other useful support features
(e.g., sorting, searching) for interacting with the data.
Multi-dimensional STL vectors are used as storage containers for
pointers to zone objects at each refinement level, which include
all the field, cell and connectivity data as class member attributes.
STL maps provide an efficient mechanism to identify direct and inverse
relations between zone iterators and unique global zone identifications.

\begin{figure}[htb]
\plotone{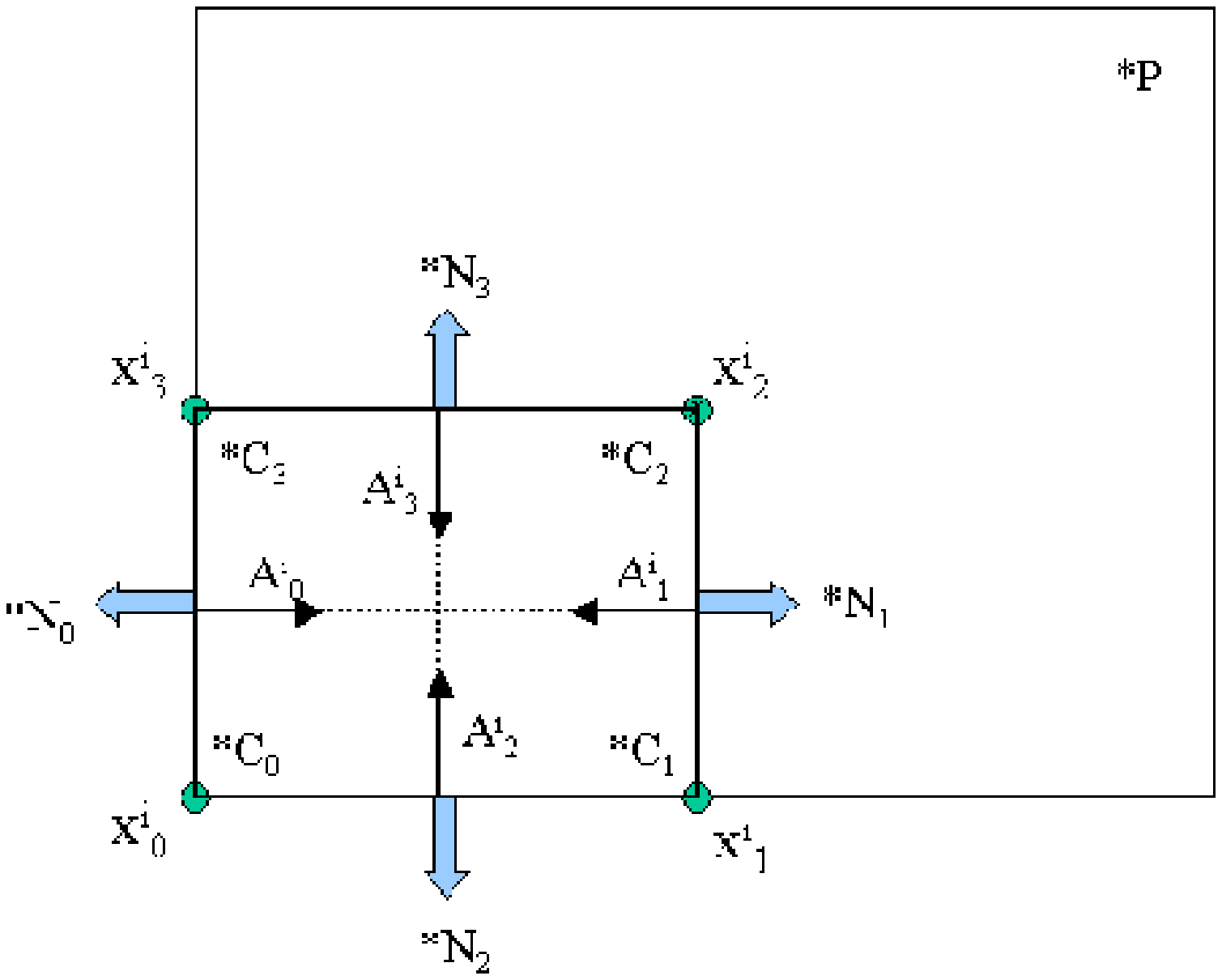}
\caption{Compilation of the local attributes stored in each cell object
to specify the geometry and connectivity data. Each cell carries
dynamically allocated pointers to its parent cell (*p),
four (eight) children (*$C_n$),
and four (six) neighbors (*$N_n$) in 2D (3D). In order to accommodate
arbitrary cell shapes and geometries, the node positions $x^i_n$, face
area normals $A^i_n$ (centered at each cell face and directed toward
the cell center), and cell volumes
are also stored and allowed to change during the evolution.
The child and neighbor lists are ordered as shown in all the
zone objects for fast and easy memory access. This data structure
allows for non-orthogonal, arbitrarily distorted and unstructured
evolving meshes.
}
\label{fig:newfig2}
\end{figure}

The essential cell geometry used in constructing meshes for \Cosmospp is
a quadrilateral (hexahedral) shape in two (three) spatial dimensions.
In order to provide enough flexibility so individual cells can
be arranged in an arbitrarily distorted and unstructured fashion,
allowing even reduced or enhanced nodal connectivities, we
store a number of attributes for each cell, including
node positions, inward (towards the cell center) pointing face
area normals, and zone volumes for convenience
(see Figure \ref{fig:newfig2}). Each zone also carries
pointers to all of its neighbors sharing a common cell face. Where a
neighbor cell has been refined to a higher level, the pointer
references the neighbor's parent so that neighbors
are always at or below the refinement level of the referencing cell
and there is at most one neighbor for each cell face.
The actual operational neighbors are found by selecting the
neighbor's appropriate children if they exist. Neighbor searches are
carried out each time a refinement cycle adjusts the mesh, and
performed with local tree scans by going up to the parent, across to the
neighbors, then down through the child zone lists. This is done to allow greater
flexibility and simplicity during the refinement cycle in case, for example,
a cell and its neighbors are tagged for de-refinement simultaneously.
When a cell is refined, it is decomposed into two, four or eight sub-cells
in one, two or three dimensions, and  pointers to these newly allocated
child zones are stored in ordered contiguous fashion for fast and easy reference.
Each child zone, in turn, carries a pointer to its parent cell.
As a result, all cells in the mesh are fully threaded with local
binary-, quad-, or oct-tree parent-child hierarchies and semi-direct
neighbor access.

\Cosmospp currently supports several refinement criteria, including
field value, normalized field gradient, normalized field curvature,
mass, and Jeans mass. When either of the field, gradient, or curvature
options are selected, the user can specify any stored variable upon
which to apply the criterion. Also, as an option, the refinement
criterion can be computed for either the
evolved fields or their conformal counterparts (by dividing out the
metric determinant), which is useful to track only flow (not metric)
features. Fields in newly created refined cells are reconstructed
by a linear, conservative, and monotonic interpolation of nearest
neighbor and parent data.

Numerical calculations are performed in the physics packages only across
the list of leaf zones which have no children. The effective mesh is thus
unstructured in general, composed of many different sized zones
containing any number of hanging nodes with reduced nodal connectivity.
In order to accomodate these kinds of meshes (and even more general ones)
we use finite volume, in place of finite difference, methods to solve the
GRMHD differential equations. Simple derivatives are computed using standard
second order (on uniform orthogonal meshes) finite volume discretization.
Gradients of a generic vector field $T_i$, for example, are evaluated
by averaging the gradient function over a single cell control volume
and converting the volume integral into a surface summation
\begin{equation}
G_{ij} \equiv \partial_j T_i =  \frac{1}{V_z} \int_{\mathrm{d}V} \partial_j T_i~\mathrm{d}V
       = -\frac{1}{V_z} \oint T_i~\mathrm{d}A_j
       = -\frac{1}{V_z} \sum_f^{\text{faces}} (T_i^*~A_j)_{f}  ~,
\label{eqn:fv_grad}
\end{equation}
where the summation is performed over all cell faces.
Here $V_z$ is the zone volume, $A_j$ is the area vector normal to
the cell face ($f$) pointing inwards toward the cell center,
$T_i^*$ is the appropriately averaged (or upwinded) field value
at the cell faces, and the gradient is returned as a tensor with
rows (columns) representing the vector (gradient) directions
(in usual gradient index notation $G_{ij} = T_{i,j}$).
The negative sign in (\ref{eqn:fv_grad}) is a result of
choosing an inward pointing area vector. At zone interfaces separating
regions of different refinement levels, we use volume weighted
averages of fields to define the ``downwind'' component for interpolation
to the cell faces.

Many of the gradient calculations appearing as source terms in a typical
time sequence update are computed using a variation of the above expression.
Some of the other, more specialized, operations used in solving the
GRMHD equations in the different formalisms are described below.

\subsection{AV Method}
\label{subsec:methods_ie}

\subsubsection{Advection}
\label{subsubsec:advection}

Advection is solved for each evolved field quantity using a time-explicit,
first order forward Euler scheme together with the scalar divergence
form of equation (\ref{eqn:fv_grad}). Letting ${\bf F}$ represent any of
the evolved fields ($D$, $E$, $S_j$, $B^j$), the discrete, finite-volume
representation of the transport source follows from (\ref{eqn:fv_grad})
\begin{equation}
\partial_i ({\bf F} V^i)
          = -\frac{1}{V_z} \sum_f^{faces} ({\bf F}^*~V^i~A_i)_f  ~,
\label{eqn:fv_adv}
\end{equation}
where $V_z$ is the local donor zone volume,
$(A_i)_f$ is the inward pointing area normal vector of face $f$,
and $(V^i)_f$ is the face-centered velocity defined as a weighted average across
neighboring cells. The quantity $({\bf F}^*)_f$ represents first order zone-centered fields
estimated at each cell face by a Taylor's expansion using limited gradient extrapolants,
${\bf F}^* = {\bf F}_{z} + (\partial_i {\bf F})^L_{z} (r^i - r^i_{z})$,
projected from the donor cell center $r^i_{z}$ to either the face center $r^i = r^i_f$
or the advection control volume $r^i = r^i_f - (\Delta t/2) (V^i)_f$, over
a time-step interval $\Delta t$.

\begin{figure}[htb]
\plotone{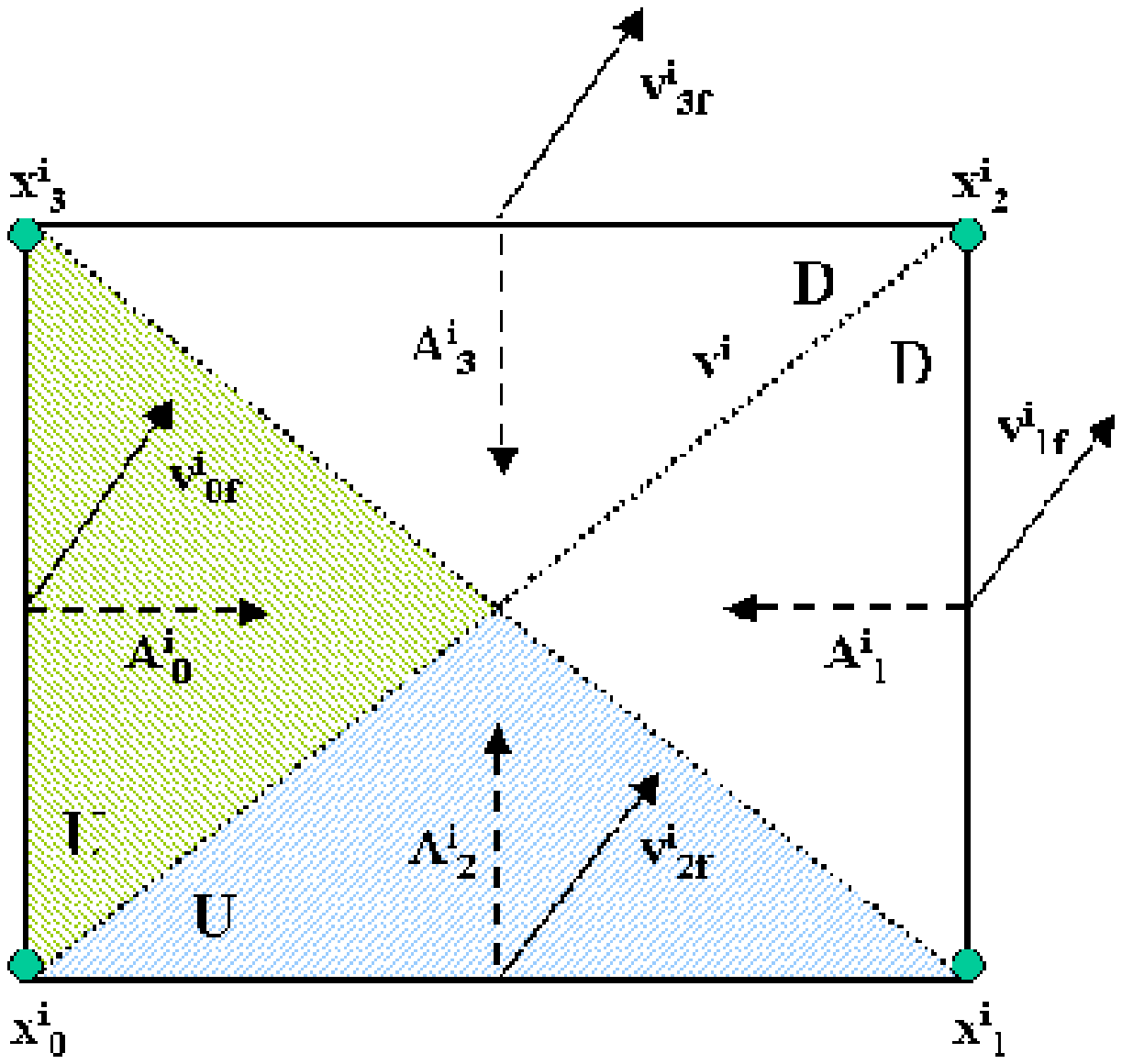}
\caption{Schematic of the sub-zonal decomposition used in constructing
upwind (U) and downwind (D) stencil geometries for enforcing
monotonicity in the gradient extrapolants. In this example, the
two shaded triangular domains comprise the total upwind polygonal sub-zone as
determined by the sign of the inner product of the face area normal
and the face-centered velocity vectors ($A_i v^i > 0$).
The two unfilled triangular domains separated by a dotted line represent
the composite downwind polygonal. Gradients are computed separately on
the three different cell blocks (upwind, downwind, and average which includes
the entire original quadrilateral cell) using finite volume contour integrals
with fields appropriately averaged to each of the cell faces.
The final gradient is a combination of the
three sub-zonal constructs and limited according to the procedures
discussed in the text for monotonicity.
In 3D, the sub-zone domains are constructed from tetrahedral elements.
}
\label{fig:newfig3}
\end{figure}

The zone-centered limited gradient $(\partial_i {\bf F})^L_z$
is constrained to force monotonicity in the extrapolated fields. In the case
of unstructured meshes, this is achieved by identifying three unique control volumes
which we assign as upstream, downstream, or average. The average control volume
is simply the total cell volume. As shown in Figure \ref{fig:newfig3},
upstream and downstream volumes are constructed
from the sub-zonal tetrahedral (triangular) in 3D (2D) geometric elements
composed of node positions connecting a cell face center, two (one) zone nodes
in 3D (2D), and the zone center. A sub-zone associated with the
cell face $f$ is defined as upstream (downstream)
if the sign of the vector product $(A_i V^i)_f$ is positive (negative).
The total upstream and downstream volumes are the sum of the tetrahedral
(triangular) sub-zones matching the corresponding signature criteria. Gradient operators
are then computed with (\ref{eqn:fv_grad}) on each of these unique,
arbitrary polyhedral (polygonal) sub-volumes, as surface integrals
with appropriately averaged fields at each surface element boundary. We denote these
gradients as $\partial^{U}_i{\bf F}$, $\partial^{D}_i {\bf F}$, and
$\partial^{A}_i{\bf F}$ for the upstream, downstream, and average, respectively.
To enforce monotonicity, the actual limited gradient $(\partial_i {\bf F})^L_z$
is set to zero if the vector product of any combination of the three gradient
operators is negative. The gradient is limited further to various degrees of
sharpness by defining the normalized scalar
$\theta_\ell = \partial^{U}_\ell {\bf F} /
               (\partial^{D}_\ell {\bf F} + \epsilon)$,
for each direction $\ell$ and smallness parameter $\epsilon \ll 1$,
and applying either of the minmod $\phi_\ell = \max[0, ~\min(1,~\theta_\ell)]$,
van Leer $\phi_\ell = (|\theta_\ell| + \theta_\ell)/(1 + |\theta_\ell|)$,
or superbee
$\phi_\ell = \max [\min(1,~2\theta_\ell),~\min(2,~\theta_\ell)]$ limiters
\citep{lev92},
to derive the final expression for the zone-centered limited gradient components
$(\partial_\ell {\bf F})^L_z = \phi_\ell \partial_\ell^D {\bf F}$.

An alternative, though somewhat more restrictive and costly, method that we
have developed for applying unstructured grid limiters on gradient
functions is based on modifying the magnitude of the average gradient with
some function of the maximum
($\partial_{\max} \equiv \max[|\partial^{U}_i|,~|\partial^{D}_i|,~ |\partial^{A}_i|]$) and
minimum ($\partial_{\min} \equiv \min[|\partial^{U}_i|,~|\partial^{D}_i|,~|\partial^{A}_i|]$)
of the three sub-zonal gradient magnitudes. For example,
\begin{equation}
(\partial_i {\bf F})^L_z = \alpha~h\left(\beta\partial_{\max}, ~(1-\beta)\partial_{\min}\right)
            \frac{\partial_i^A {\bf F}}{|\partial_i^A {\bf F}| + \epsilon} ~,
\end{equation}
where $\beta$ is a steepness parameter bounded by $0\le \beta \le 1$,
$h( ... )$ is any function of the arguments,
and $\alpha$ is a coefficient to enforce monotonicity in the
extrapolated field ${\bf F}^*_f$. In particular, we set
$\alpha = \min(1,~\max(0,~\min(\alpha_1,~\alpha_2)))$, where
\begin{equation}
\alpha_1 =
\begin{cases}
      \frac{\max(0,~{\bf F}_{z,n} - {\bf F}_z)}{{\bf F}_f - {\bf F}_z}
      & \text{if}\quad
        {\bf F}_f = {\bf F}_z + (\partial_i {\bf F})^L_z (r^i_f - r^i_z)/\alpha
        > \max({\bf F}_z, {\bf F}_{z,n})    \\
      1
      & \text{otherwise}
\end{cases}
\label{eqn:alpha_1}
  ~,
\end{equation}
and
\begin{equation}
\alpha_2 =
\begin{cases}
      \frac{\min(0,~{\bf F}_{z,n} - {\bf F}_z)}{{\bf F}_{f} - {\bf F}_z}
      & \text{if}\quad
        {\bf F}_{f} < \min({\bf F}_z, {\bf F}_{z,n}) \\
      1
      & \text{otherwise}
\end{cases}
\label{eqn:alpha_2}
  ~,
\end{equation}
where ${\bf F}_{z,n}$ refers to the field value in the neighbor zone center.
The min/max operations in (\ref{eqn:alpha_1}) and (\ref{eqn:alpha_2})
are performed over each of the faces in the donor cell, and $\alpha$ is chosen
as the smallest value needed for strict monotonicity across all cell faces.
Of course these procedures for solving the advection equation and evaluating limited
gradients reduce in one spatial dimension to the familiar upwind scheme as described,
for example, in \cite{haw84b}.

\subsubsection{Artificial Viscosity}
\label{subsubsec:artificial}

We have implemented five different artificial viscosity options for shock
capturing. All of these constructs have a common inertia
multiplier defined for relativistic MHD as
\begin{equation}
I = D + E + W(P + ||Q|| + 2 P_B) ~.
\label{eqn:inertia}
\end{equation}
The inertia is also normalized by a factor $I_F$ that scales out the local 3-geometry
in curvilinear coordinates, and introduces a relativistic boost factor
to an arbitrary power that is useful for effectively transforming the length
scale between proper and boosted frames, and providing a flexible scaling
with boost. The normalized inertia multiplier is thus defined as
\begin{equation}
I_N = \frac{I}{I_F} = \left(D + E + W(P + ||Q|| + 2 P_B\right)
                      \frac{1}{\sqrt{\gamma}} \left(\frac{\sqrt{-g}}{W}\right)^n ~,
\end{equation}
where $\gamma$ is the 3-metric determinant, and $n$ is an arbitrarily specifiable constant.

The simplest artificial viscosity we consider is the scalar form based on the work of
\cite{von50}:
\begin{equation}
Q^i_j =
\begin{cases}
      I_N~\Delta l~\partial_k V^k~(k_q~\Delta l~\partial_k V^k - k_l~C_s) \delta^i_j
      & \text{if}\quad
        \partial_i V^i < 0   \\
      0
      & \text{otherwise}
\end{cases}
\label{eqn:vis_scalar}
  ~,
\end{equation}
where $\Delta l$ is the minimum covariant
zone length, $C_s$ is the local sound speed, and $k_q$ and $k_l$ are constant coefficients
multiplying the quadratic and linear contributions, respectively. This scalar viscosity
remains one of the most popular representations due to its simplicity
and robustness; however, it is also the most diffusive of the methods
we have implemented since it
indescriminantly filters out transverse, longitudinal, and sheared
isotropically averaged compressive flows.

A second option for viscosity is similar to (\ref{eqn:vis_scalar})
but is designed for anisotropic
shock capturing along the principal grid axes similar to what is
achieved in directionally split,
fully structured mesh approaches:
\begin{equation}
Q^i_j = \frac{I_N~\Delta l}{2} (k_q~\Delta l~\partial_i V^i - k_l~C_s)
        ~\text{diag} \left[\partial_x V^x - |\partial_x V^x|, ~
                           \partial_y V^y - |\partial_y V^y|, ~
                           \partial_z V^z - |\partial_z V^z|
                     \right]~,
\end{equation}
where $\text{diag}[ ... ]$ refers to a list of the main diagonal elements of a tensor.
This is generally less diffusive than the isotropic scalar viscosity option
since it is automatically disengaged along any grid direction that is
not undergoing compression.

Another option is based on the scalar viscosity developed by \cite{whi73}, but
extended here for relativistic problems. Defining the quantity
\begin{equation}
\widetilde{Q} = \Delta l~|\partial_i V^i|^{1/2}~
            |g^{ij} \partial_i P~\partial_j (I_N^{\ -1})|^{1/4} ~,
\end{equation}
this viscosity can be written as
\begin{equation}
Q^i_j = I_N~\widetilde{Q}~(k_q~\widetilde{Q} + k_l~C_s) \delta^i_j
\label{eqn:vis_white}
\end{equation}
when both conditions, $g^{ij} \partial_i P~\partial_j (I_N^{\ -1}) < 0$ and
$\partial_i V^i < 0$, are met. The advantage of (\ref{eqn:vis_white}) is that
both pressure and velocity gradients are incorporated into its definition,
which is a better indication of the presence of shocks than the
velocity gradient by itself. It can therefore be more effective at
suppressing artificial heating in regions undergoing adiabatic compression.

The final two options we consider are genuine tensor viscosities.
The first of these is similar to \cite{tsc79}, but does not include the
full covariant gradient treatment. However, since the velocity gradients
are computed in the conformal frame and the proper volume is accounted for
in the inertia normalization, the version we have implemented
\begin{equation}
Q^i_j = I_N~\Delta l~(k_q~\Delta l~\partial_i V^i - k_l~C_s)
        ~\text{Sym}\left(\partial_j V^i - \frac{\partial_i V^i}{3} \delta^i_j\right)
        ~,
\end{equation}
where $\text{Sym}( ... )$ denotes a symmetry operation,
is a reasonable simplification. This traceless tensor viscosity generally
outperforms scalar viscosities in preserving geometric symmetries (e.g., sphericity),
and suppressing viscous heating in homologous spherical contraction.

The last option we consider is potentially the least diffusive, but also the most
unstable or unpredictable method due to the matrix inversion operations needed to
align the viscosity calculation to the natural principal axes of the shock frame.
The shock orientation and velocity differences are determined by the eigenvectors
$R^i_k(S)$ and eigenvalues $\lambda^i(S)$ of the symmetrized strain rate tensor
$S^i_j = \text{Sym} (\partial_j V^i - \partial_i V^i~\delta^i_j/3)$.
In this procedure the artificial viscosity is written as
\begin{equation}
Q^i_j = R^i_k(S)~\widehat{Q}^k_l~R^l_j(S)^T ~,
\end{equation}
where $R^l_j(S)^T$ is the transpose of $R^i_k(S)$, $\widehat{Q}^k_l$ is
the usual von Neumann-Richtmyer viscosity in the shock aligned frame
where the viscosity tensor has only diagonal elements
\begin{equation}
\widehat{Q}^i_i = I_N~\widehat\Delta l^i~\lambda^i(S)~
              \left(k_q~\widehat\Delta l^i~\lambda^i(S) - k_l~C_s\right) ~.
\end{equation}
The shock-aligned length scales $\widehat\Delta l$ are computed in a similar
fashion, also accounting for local proper frame metric distortions,
$M_{ij} = g_{ij} \Delta x^i \Delta x^j$ (assuming no index summation here).
Representing the eigenvalues and eigenvectors of the matrix $M_{ij}$ as
$\lambda^i(M)$ and $R_{ij}(M)$, we define the shock-aligned length scales as
the diagonal elements of the transformed length scale tensor
$\widehat\Delta l^i = \text{diag}[\widehat\Delta l_{ij}]$, where
\begin{equation}
\widehat\Delta l_{ij} = R_{ia}(S)^T ~R_{ab}(M) ~\sqrt{\lambda^{bc}(M)}
                       ~R_{cd}(M)^T ~R_{dj}(S) ~,
\end{equation}
and $\lambda^{bc}(M)$ is the diagonal tensor of eigenvalues $\lambda^i(M)$.
This expression effectively transforms the principal length scales
$\lambda^i(M)$ to the grid axes, then to the principal frame of
the strain rate tensor.

Because the basic strain rate tensor (and especially velocity divergence)
does not always distinguish between discontinuous shocks and smooth compressive
flows, it is sometimes necessary to be more selective about where artificial
viscosity is applied. As an option we can modify the strain rate tensor as
\begin{equation}
\partial_i V^j \Rightarrow \partial_i V^j - k_L~\partial_i^{L} V^j ~,
\end{equation}
where $k_L$ is a constant less than unity (typically 0.5), and
$\partial_i^{L}$ is a limited gradient. This effectively reduces the
viscosity levels over smooth flows where the limiter has no affect, while
keeping it strongly active over shocks where the limited gradient vanishes.
In this case we use a hard limiter and set $\partial_i^{L} V^j = 0$ if
any of the velocity or velocity gradient components have opposite signs across
any of the nearest neighbor zones. If both the velocity and its gradient
are monotonic in the local neighbor patch, we set
$\partial_i^{L} V^j = \partial_i V^j$.

\subsection{NOCD Method}
\label{subsec:nocd}

We take a slightly different and significantly simpler
approach in this paper to solving the fully
conservative system of equations described in \S\ref{sec:conservative_e}
than the Riemann-free method we implemented in \Cosmos \citep{ann03a}.
This approach, developed by \cite{kur00}, is also part of the NOCD family of
solutions, but we find it to be generally less diffusive and less
sensitive to Courant restrictions.

The curvature sources in equations (\ref{eqn:hr_en}) and (\ref{eqn:hr_mom})
are updated using finite volume discretization,
and a first order in time Euler advance method.
The divergence terms in (\ref{eqn:hr_en}), (\ref{eqn:hr_mom})
and (\ref{eqn:hr_ind}) are updated in time from $u^n$ to $u^{n+1}$
using a general $N$th order
method consisting of a sequence of first order Euler steps \citep{shu88}.
Writing the time update in terms of forward Euler templates,
high order time solutions can be expressed as
\begin{eqnarray}
u^{(1)}   &=& u^n + \Delta t^n S(u^n) ~, \nonumber \\
u^{(m+1)} &=& \eta_m u^n + (1-\eta_m)(u^{(m)} + \Delta t^n~S(u^{(m)})) ~, \\
u^{n+1}   &=& u^k ~, \nonumber
\end{eqnarray}
for ordered sequences $m=1,~2,~...,~k-1$ up to $k$ steps, an arbitrary source term $S(u)$,
and constant parameters
\begin{equation}
\eta =
\begin{cases}
      (1/2,~-)~,   \quad & \text{$2$nd order} \\
      (3/4,~1/3)~, \quad & \text{$3$rd order}
\end{cases}
~.
\end{equation}
At first order, $\eta_m$ is not used since that reduces to a simple,
single-step forward Euler solver. The time update is thus conveniently
solved as a sequence of first order Euler cycles, and is easily extended
to higher order using these simple prescriptions.

In semi-discrete form, the single-step solution to a general nonlinear
convection equation for a general field $\omega$,
representing the energy (\ref{eqn:hr_en}),
momentum (\ref{eqn:hr_mom}), or induction (\ref{eqn:hr_ind}) equations,
is written in finite volume form as
\begin{equation}
\partial_t \omega = -\partial_i F^i(\omega) = \frac{1}{V_z}\sum_f^{\text{faces}}
           \left[ \frac{1}{2}\left(F^i_f(\omega^-_f) + F^i_f(\omega^+_f)\right) (A_i)_f
                 -\frac{a_f}{2}\left(\omega^-_f - \omega^+_f\right) ||A_f||
           \right]
  ~,
  \label{eqn:hr_solution}
\end{equation}
where $F^i$ are the numerical fluxes,
$a_f = \max(a^+_f, ~a^-_f)$ is the maximum local propagation speed,
and $\omega^{\pm}_f = \omega_\pm + (\partial_i \omega)^L_\pm ~(r^i_f - r^i_\pm)$
are the limited gradient projections of the cell-centered field $\omega_\pm$
to the cell faces.
The subscripts $\pm$ refer to either the donor cell center ($-$) or the opposite
zone ($+$) across face $f$, and the superscripts $\pm$ refer to the face centered
projections of the field originating from the corresponding zone. In this approach,
the transport terms are grouped together with the other fluxes, and the
discrete solution (\ref{eqn:hr_solution}) is applied to all divergence
source terms (transport included) simultaneously.

\subsection{Extended AV Method}
\label{subsec:eav}

The basic internal energy formulation with artificial viscosity can be
expanded easily to include an additional equation for the conserved energy
in the form (\ref{eqn:hr_en}).
The general idea in this dual energy approach is to extract the thermal component
from the total energy, and depending on the accuracy of the result,
use it to over-write
the solution computed directly from the internal energy evolution equation.
However, care must be used when extracting the internal energy, since
it is often the case that adiabats in total energy conserving methods
can be grossly miscalculated.

In this scheme, the total energy is used as an option to compute the
four-momentum normalization
at various stages of the solve sequence, as well as the inertia for the
artificial viscosity when certain conditions are met.
Defining ${\cal E}_D$ as the non-thermal or ``dynamical''
component of the conservative energy,
\begin{equation}
{\cal E}_D = \frac{DW}{\sqrt{-g}} + \frac{2 P_B W^2}{\sqrt{-g}}
           + \sqrt{-g}\left( g^{00} P_B - \frac{B^0 B^0}{4\pi}\right) ~,
\end{equation}
we write
\begin{equation}
\widetilde{E} = \frac{({\cal E}-{\cal E}_D)\sqrt{-g}~W}
                {\Gamma W^2 + (\Gamma-1)g^{00}(\sqrt{-g})^2}  ~,
\end{equation}
for the internal energy extracted from the conserved energy field,
and
\begin{eqnarray}
I &=& W(\rho h + 2P_B) = D + E + W(P + ||Q|| + 2P_B) \nonumber \\
  &=& \frac{\sqrt{-g}}{W}\left( {\cal E} - \sqrt{-g} g^{00}(P + P_B)
                                       + \frac{1}{4\pi}\sqrt{-g} B^0 B^0\right)
\end{eqnarray}
for the inertia and momentum normalization.

Also, an additional term is added to the total energy evolution equation that
accounts for collisional dissipation heating arising from the artificial viscosity.
The conservative energy equation that we solve takes the general form
\begin{equation}
\partial_t {\cal E} + \partial_i \left({\cal E}V^i\right)
                    + \partial_i \left(F^i\right)
            = \Sigma^0 ~,
\label{eqn:dual_en}
\end{equation}
where the flux $F^i$ is now defined as
\begin{equation}
   F^i = \sqrt{-g}~\left( (g^{0j} - g^{00} V^j)~((P+P_B) \delta^i_j + Q^i_j)
       - \frac{1}{4\pi}(B^i B^0 - B^0 B^0 V^i) \right)
       ~.
\end{equation}
Although we use essentially the same equation for energy
(apart from $Q^i_j$) as the NOCD
method described in the previous section, we solve it using a different,
more conventional operator split approach. Here we solve sequentially for
curvature effects ($\partial_t {\cal E} = \Sigma^0$), followed by the
transport source ($\partial_t {\cal E} + \partial_i ({\cal E} V^i) = 0$)
which is updated synchronously with the momentum, internal energy, and magnetic
induction advection, then finally we apply second order finite volume discretization
to evaluate the remaining flux divergence term
($\partial_t {\cal E} + \partial_i F^i = 0$).

The final step in this extended scheme is to determine whether the total energy
solution is known well enough to extract accurately the internal energy.
We use for a measure of accuracy, the minimal difference ratio
\begin{equation}
\left(\frac{\delta {\cal E}}{\cal E}\right)_{\min} \equiv
   \frac{\text{min}[ E~h(\Gamma),
      ~\text{max}(\rho\epsilon_{\text{floor}} W h(\Gamma),
      ~\widetilde{E}~h(\Gamma))]_N}
   {\text{max}({\cal E})_N}
~,
\end{equation}
where
\begin{equation}
h(\Gamma) = \frac{\Gamma W^2 + (\Gamma-1) g^{00} (\sqrt{-g})^2}
                 {\sqrt{-g}~W} ~,
\end{equation}
$\rho\epsilon_{\text{floor}}$ is the minimum allowable energy density threshold,
and the subscript $N$ refers to extending local minimum and maximum
calculations to include all neighbor zones, adding an extra measure of safety.
A common problem with total energy schemes in general is that numerical
truncation errors can accumulate to the point that the sum of different physical
contributions often exceeds the total energy (${\cal E}_D > {\cal E}$),
especially in kinematic or magnetic field
dominated flows, and in the vicinity of strong shocks. This problem is
avoided by forcing a minimum threshold on either $E$ or $\widetilde{E}$
to guarantee positivity, and by preserving the solution from the internal
energy equation whenever ${\cal E}_D \ge {\cal E}$, or
$(\delta{\cal E}/{\cal E})_{\text{min}} \le \delta_c$,
where $\delta_c$ is a user defined parameter large enough to prevent
numerical noise from corrupting the solution. The internal energy
is otherwise set to $E = \text{max}(\epsilon_{\text{floor}} W, ~\widetilde{E})$
when $(\delta{\cal E}/{\cal E})_{\text{min}} > \delta_c$ and
the extracted thermal component can be trusted.

\section{Code Tests}
\label{sec:tests}

\subsection{Hydrodynamics}
\label{sec:hydrotests}

\subsubsection{Shock Tube}
\label{sec:stube}

\begin{figure}[htb]
\plotone{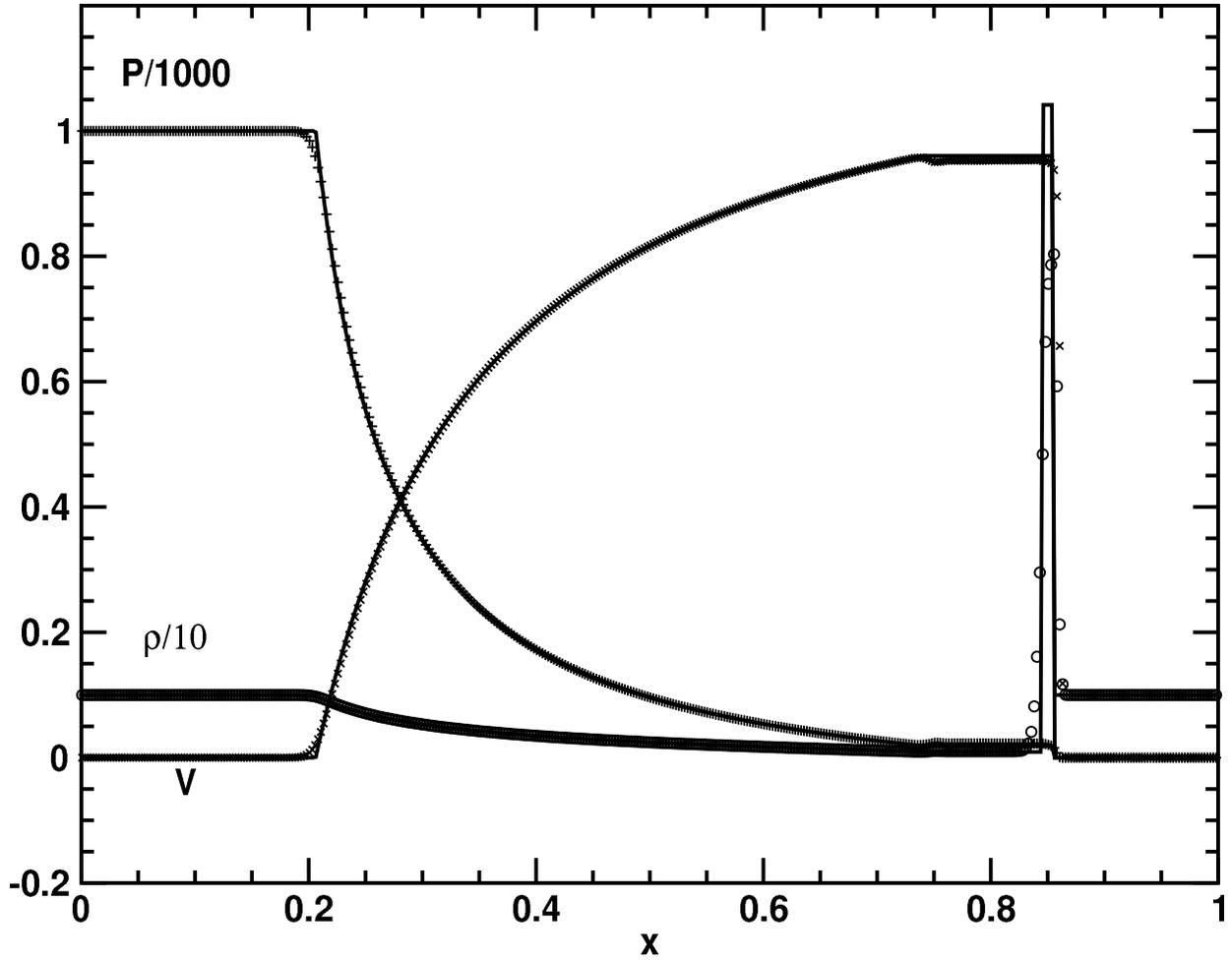}
\caption{Results at time $t=0.36$
for the shock tube test using artificial
viscosity (AV) and 800 zones.  The data points in this plot have been
sampled to reduce overcrowding.  Only 400 points are shown. The solid
line shows the analytic solution.}
\label{fig:stube_AV}
\end{figure}

\begin{figure}[htb]
\plotone{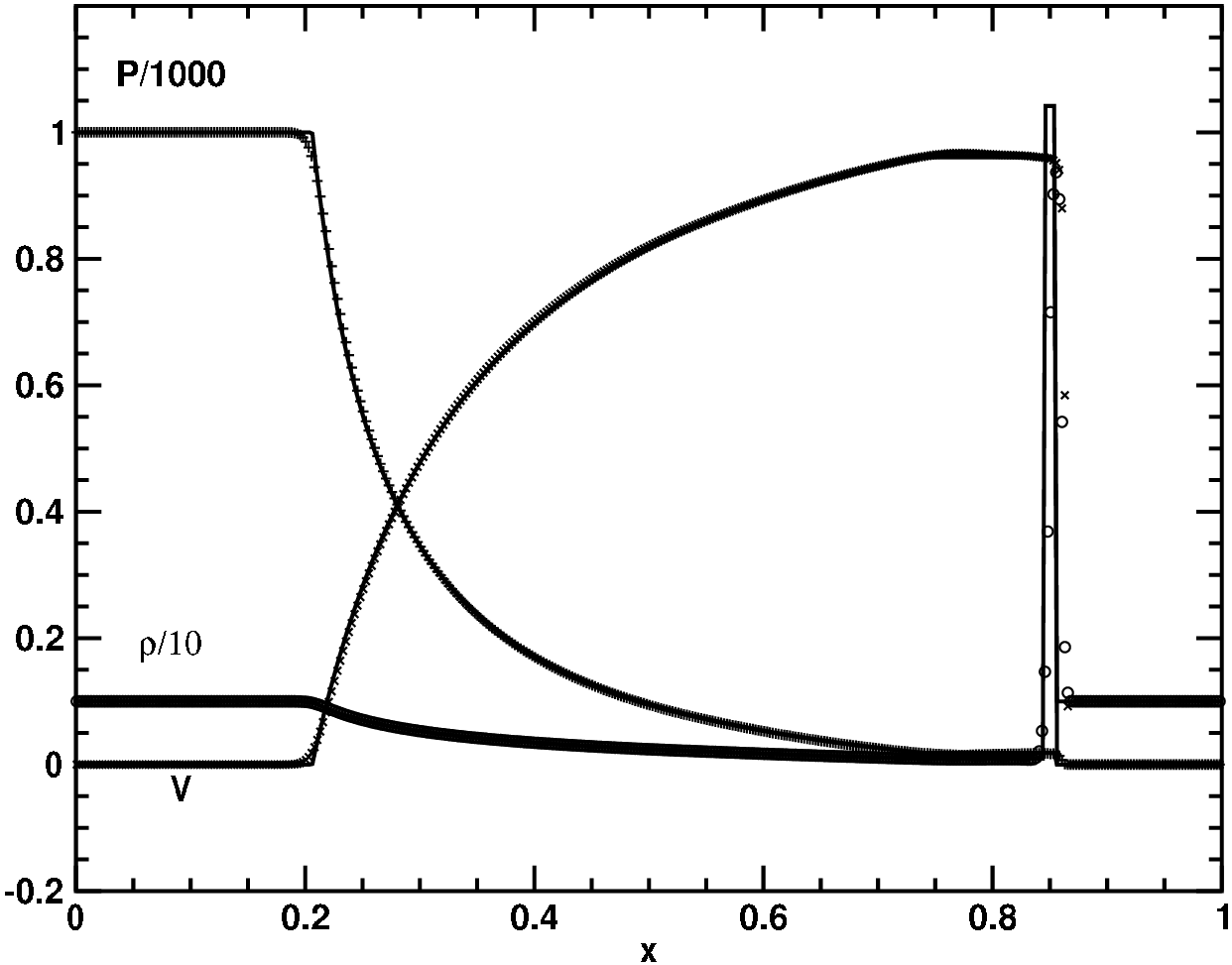}
\caption{As Figure \protect{\ref{fig:stube_AV}} but with
the extended viscosity (eAV) scheme.}
\label{fig:stube_Hyb}
\end{figure}

\begin{figure}[htb]
\plotone{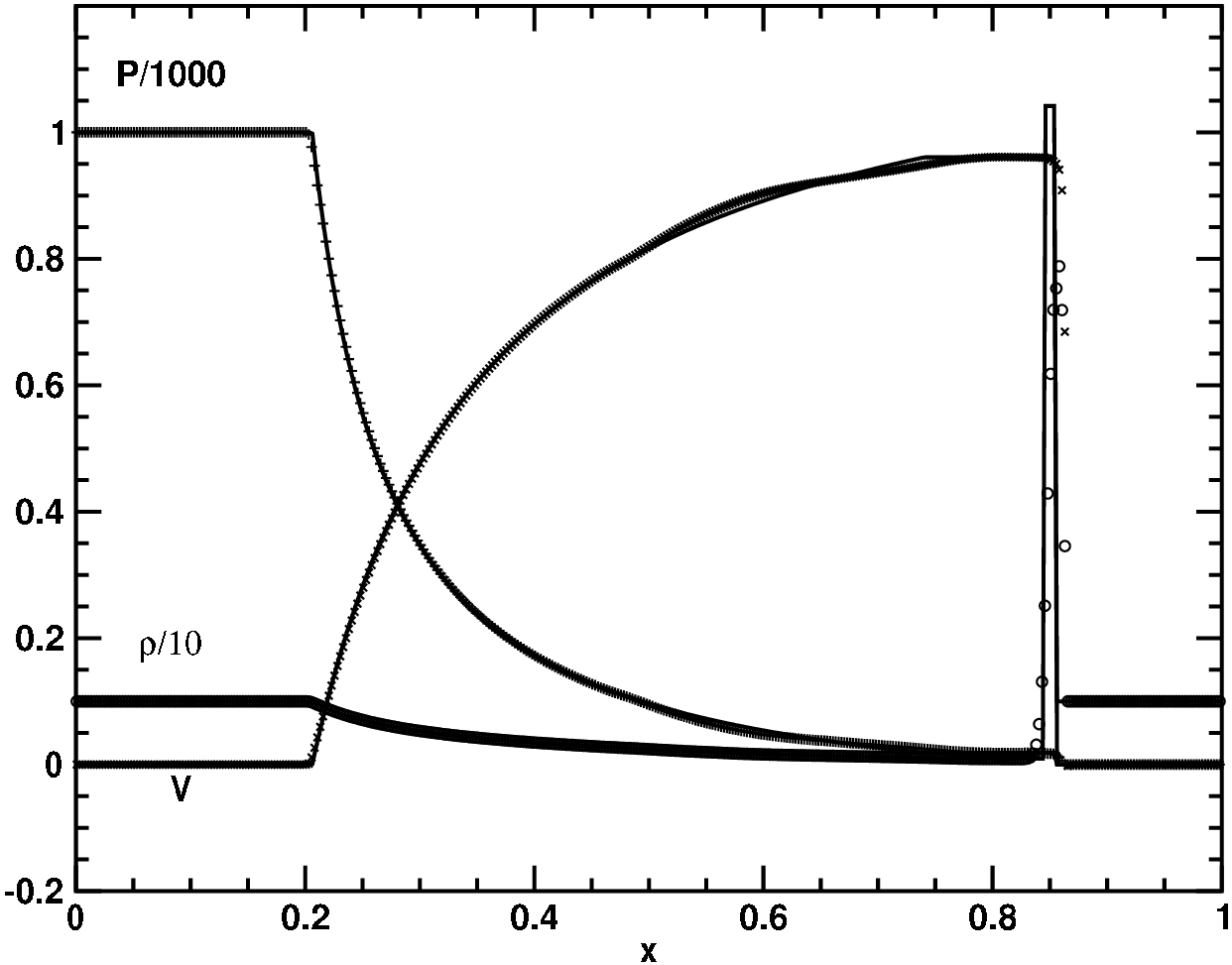}
\caption{As Figure \protect{\ref{fig:stube_AV}} but with
the non-oscillatory central difference (NOCD) scheme.}
\label{fig:stube_NOCD}
\end{figure}

We begin testing with one of the standard problems in fluid
dynamics, the shock tube, in which two different fluid states are
initially separated by a membrane. At $t=0$ the membrane is removed
and the fluid evolves in such a way that five distinct regions
appear in the flow: an undisturbed region at each end, separated by
a rarefaction wave, a contact discontinuity, and a shock wave.
Although this problem only checks the hydrodynamic elements of the
code, as it assumes a flat background metric and ignores magnetic
fields (magnetosonic shocks will be considered in \S
\ref{sec:mhdtests}), it is still useful for evaluating the
shock-capturing properties of the different methods. We consider the
high boost ($W=3.59$, $\Gamma=5/3$)
case from \citet{ann03a}. The initial state is
specified as $\rho_L = 1$, $P_L = 10^3$, $V_L = 0$ to the left of
the membrane and $\rho_R = 1$, $P_R = 10^{-2}$, $V_R = 0$ to the
right. The membrane is located at $x=0.5$ on a grid of unit length.
The results presented here are run using the scalar artificial
viscosity with a quadratic viscosity coefficient $k_q=2.0$, linear
viscosity coefficient $k_l=0.3$, Courant factor $k_{cfl}=0.3$,
viscosity multipier $k_{\dot{W}}=0$, and are carried out on fixed,
uniform grids of differing resolutions in order to establish the
convergence of each method. Figures \ref{fig:stube_AV},
\ref{fig:stube_Hyb}, and \ref{fig:stube_NOCD} show spatial profiles
of the results at time $t=0.36$ on a grid of 800 zones using the AV,
eAV, and NOCD methods, respectively. Table \ref{tab:errors1}
summarizes the errors in the primitive variables $\rho$, $P$, and
$V$ for four different grid resolutions (400, 800, 1600, and 3200
zones) and the three different CFD methods (AV, eAV, and NOCD) using
the $L$-1 norm (i.e., $\Vert E(a) \Vert_1 = \sum_{i,j,k} \Delta x_i
\Delta y_j \Delta z_k \vert a_{i,j,k}^n - A_{i,j,k}^n \vert$, where
$a_{i,j,k}^n$ and $A_{i,j,k}^n$ are the numerical and exact
solutions, respectively, and for one-dimensional problems the
orthogonal grid spacings are set to unity). All three methods
converge at approximately first order as expected for this class of
problem, and the errors are consistent with those reported for our
previous code \citep{ann03a}.

\subsubsection{Shock Collision}

A second test presented here is the wall shock problem involving
the shock heating of cold fluid hitting a wall at the left boundary
($x=0$) of a unit grid domain. The initial data are set up to be
uniform across the grid with adiabatic index $\Gamma=4/3$,
pre-shocked density $\rho_{1} = 1$,
pre-shocked pressure $P_{1} = (\Gamma -1)\times 10^{-8}$, and
velocity $V_{1} = -v_{init}$. When the fluid hits the wall
a shock forms and travels to the right, separating the initial pre-shocked
conditions from the post-shocked state
($\rho_2,~P_2,~V_2$) with solution in the wall frame
\begin{equation}
V_S  = \frac{\rho_{1} W_{1} V_{1}}{\rho_{2} - \rho_{1}W_{1}} ,
\label{eqn:vs_shock}
\end{equation}
\begin{equation}
P_{2} = \rho_{2} (\Gamma - 1)(W_{1} - 1) ,
\label{eqn:p_2shock}
\end{equation}
\begin{equation}
\rho_{2} = \rho_{1} \left[ \frac{\Gamma + 1}{\Gamma - 1} +
               \frac{\Gamma}{\Gamma - 1}(W_{1} - 1) \right] ,
\label{eqn:rho_2shock}
\end{equation}
where $V_S$ is the velocity of the shock front, and the pre-shocked
energy and post-shocked velocity are both assumed
negligible $(\epsilon_1,~ V_2) \rightarrow 0$.
All of the results in this section are performed
on a 200 zone uniformly-spaced mesh and run to a final time of
$t=2.0$. For the AV and eAV methods, we use the scalar viscosity with
$k_q = 2.0$ and $k_l = 0.7$.
The Courant factor is set to $k_{cfl} = 0.3$ for all methods.

\begin{figure}[htb]
\plotone{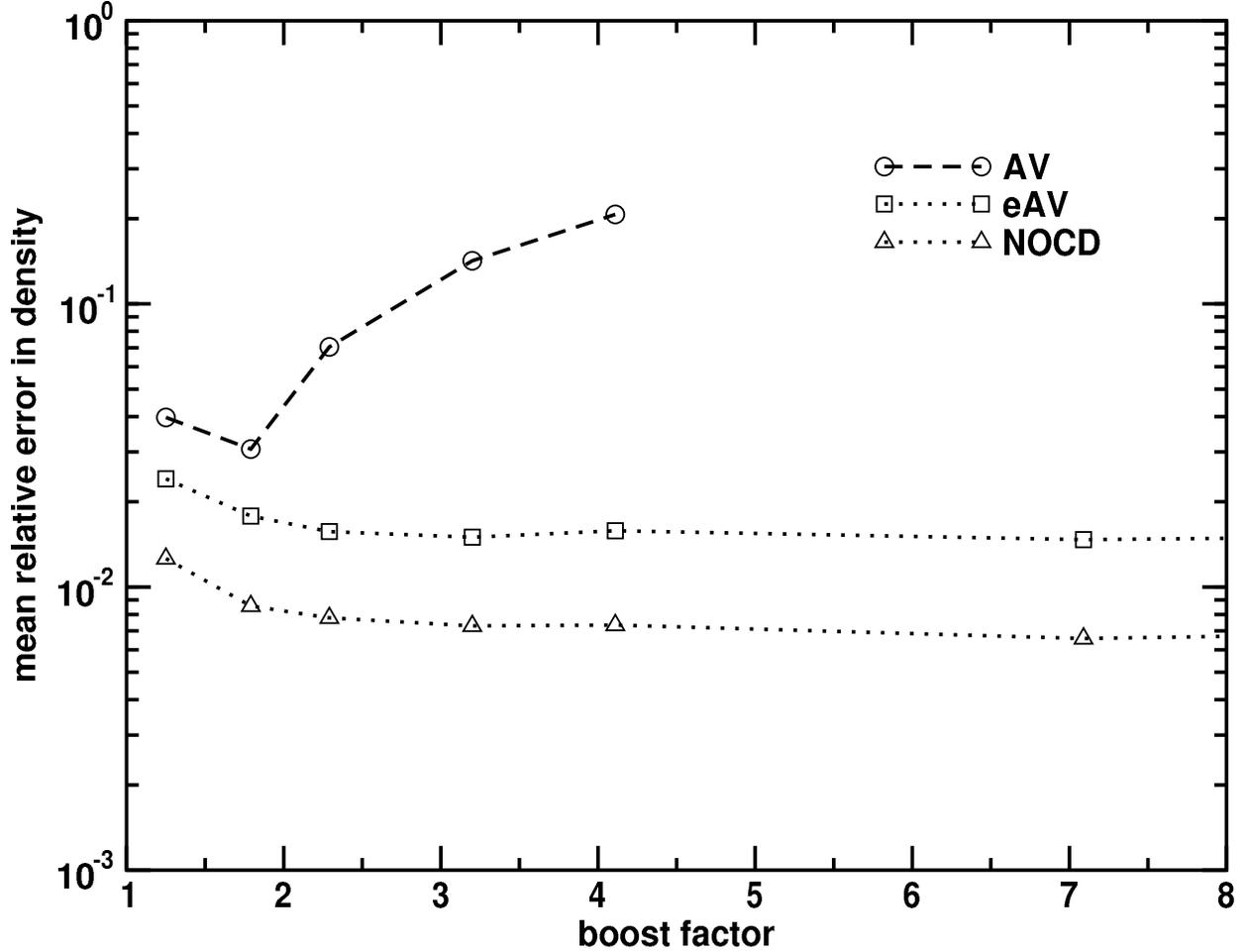}
\caption{Mean relative errors in density for the AV, eAV,
and NOCD methods as a function of boost for the relativistic wall shock
problem. All calculations were run using 200 zones up to time $t=2.0$.
The AV and eAV results can be improved significantly
and brought closer in alignment with the NOCD results
by reducing the Courant factor or increasing the viscosity
coefficients over the canonical values we have chosen.
}
\label{fig:wall_error}
\end{figure}

\begin{figure}[htb]
\plotone{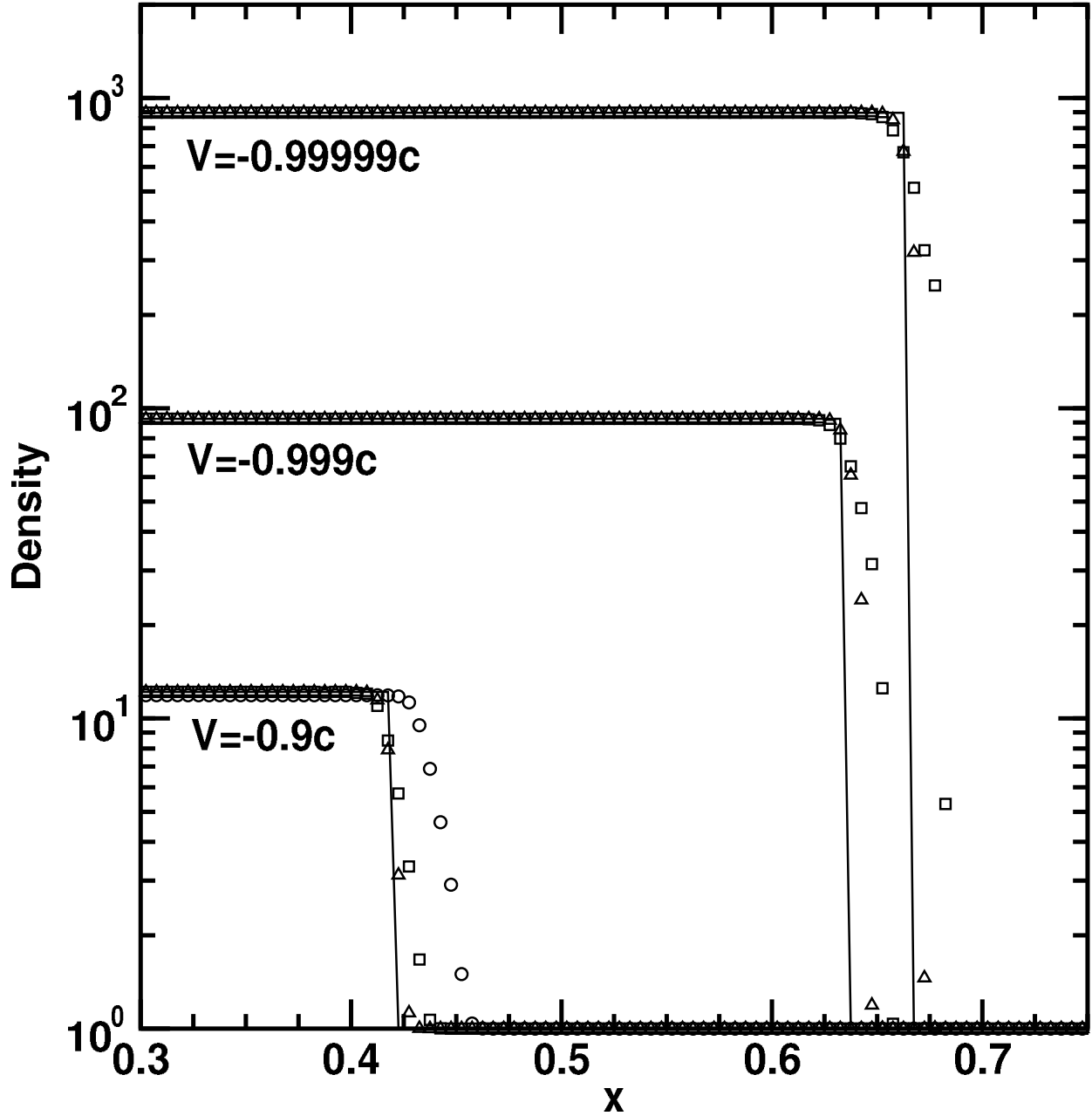}
\caption{Density plots for different infall velocities in the wall shock
test using the eAV ({\em squares}) and NOCD ({\em triangles}) methods.
Results for the AV method ({\em circles}) are only included for $V=-0.9$.
The resolution is 200 zones, the displayed time is $t=2.0$,
and the graphs are zoomed in at the shock fronts to distinguish
the different solutions.}
\label{fig:wall_den}
\end{figure}

Figure \ref{fig:wall_error} plots the mean-relative errors
in density, which are generally greater than errors in either the pressure
or velocity, as a function of boost factor.
Although we are not able to extend the AV method reliably
(which we define by a 10\% mean error threshold and increased
sensitivity to viscosity parameters)
beyond $v_{init} \sim 0.95$, the eAV and NOCD methods are
substantially more robust. As shown in
Figure \ref{fig:wall_den}, both methods can be run up to arbitrarily
high boost factors
($v_{init} > 0.99999$) with mean relative errors typically
remaining below 2\% with no significant increasing trend.
As noted previously \citep{ann03a}, the errors for the AV method
can be improved significantly by either lowering the Courant factor
or increasing the viscosity coefficients.  However, the sensitive
dependence on these parameters detracts considerably from the
attractiveness of this method for this class of problem.
The errors for the eAV method can also be improved by adjusting the
viscosity coefficients, although we find that it performs well even
with the ``standard'' values used here.

\subsubsection{Boosted Shock Collision}
\label{sec:boostcoll}

An elaboration of the wall shock problem from the previous section is the
collision of two boosted fluids.  The fluids flow toward each other and
collide, each shocking against the other and forming a contact discontinuity.
In the center-of-momentum frame, in which the contact discontinuity between
the fluids is stationary, this system is equivalent to a pair of opposing wall
shocks, each impinging on the other.  However, by boosting this system's
center-of-momentum frame with respect to the grid frame, one devises a very
rigorous test of the Lorentz invariance of the code under nonsymmetric
conditions, with multiple jump discontinuities and highly relativistic shock
velocities.

In the center-of-momentum (primed) frame, a fluid with initial state
($\rho_1,~P_1,~V'_1 > 0$) flows from the left boundary while a fluid
of state ($\rho_4,~P_4,~V'_4 < 0$) flows from the right; we assume each
fluid is cold: $P_1 = P_4 = 0$. Upon collision, fluid 1 is shocked
into a state ($\rho_2,~P_2,~V'_2$) while fluid 4 is shocked into state
($\rho_3,~P_3,~V'_3$), where the fluids are numbered sequentially from
left to right.  In this frame, each shocked fluid comes to rest; $V'_2
= V'_3 = 0$. This implies force equilibrium between these fluids; $P_2
= P_3$.  Using equations (\ref{eqn:p_2shock}) and
(\ref{eqn:rho_2shock}), pressure balance gives
\begin{equation}
  \rho_1 (\Gamma W'_1 + 1)(W'_1 - 1) = \rho_4 (\Gamma W'_4 + 1)(W'_4 - 1) ~.
\label{eqn:P2eP3}
\end{equation}
The proper densities of the shocked fuids (from eqn.~\ref{eqn:rho_2shock})
are
\begin{eqnarray}
    \rho_{2} &=& \rho_{1} \frac{\Gamma W'_1 + 1}{\Gamma - 1} \\
    \rho_{3} &=& \rho_{4} \frac{\Gamma W'_4 + 1}{\Gamma - 1} ~,
\label{eqn:rho_boostshock}
\end{eqnarray}
and the specific energies are \citep[e.g.][]{haw84b}
\begin{eqnarray}
  \epsilon_2 &=& W'_1 - 1 \\
  \epsilon_3 &=& W'_4 - 1 ~.
\label{eqn:eps_boostshock}
\end{eqnarray}
Pressure balance again implies $\rho_2 \epsilon_2 = \rho_3
\epsilon_3$. Also, from equation (\ref{eqn:vs_shock}) the reverse shock
velocity, $V'_r$, between materials 1 and 2, and the forward shock velocity,
$V'_f$, between fluids 3 and 4 are expressed as
\begin{eqnarray}
  V'_r &=& - \frac{(W'_1 - 1)(\Gamma - 1)}{W'_1 V'_1} \\
  V'_f &=& - \frac{(W'_4 - 1)(\Gamma - 1)}{W'_4 V'_4} ~.
\label{eqn:shockfronts}
\end{eqnarray}
Typically the grid (lab frame) velocities $V_1$ and $V_4$ are known and the
velocity of the center-of-momentum frame and contact discontinuity, $V_d$,
must be solved for numerically by substituting the boost transformations
\begin{equation}
  W'_1 = W_d W_1 (1 - V_1 V_d)~, \quad  W'_4 = W_d W_4 (1 + V_4 V_d)
\label{eqn:boostxforms}
\end{equation}
into equation (\ref{eqn:P2eP3}), where $W_i = (1 - V_i^2)^{-1/2}$.  The system
can also be boosted into a desired frame by standard velocity addition.

\begin{figure}
\includegraphics[width=4.5in, angle=-90]{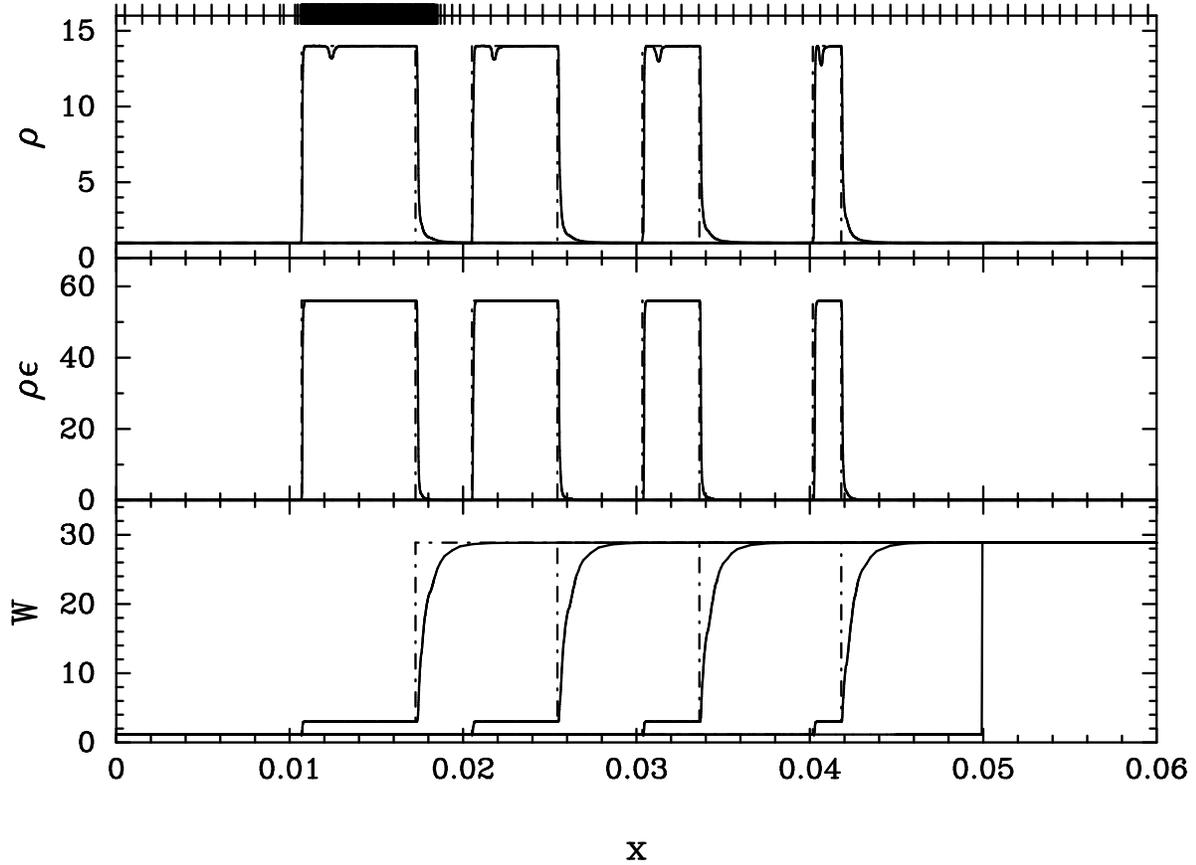}
\caption{Two ideal fluids, each with initial proper density $\rho =
1$, proper energy $\rho\epsilon = 10^{-8}$, and $\Gamma = 5/3$,
collide initially at $x = 0.05$.  The fluids move in opposite
directions, each with $W' = 5$ in the center-of-momentum frame.  The
observer is boosted to the right at $W = 3$ so the fluid moves at
speeds up to $V/c \sim 0.999$, and the shocked fluid region can be
seen to move to the left over a sequence of five times, $t = $ 0.0,
0.01, 0.02, 0.03, 0.04.  The numerical results are solid lines and
show good agreement with the analytical solution (Table
\ref{tab:boostcol}) shown in dot-dashed lines.  The ticks at the top
of the plot are zone positions for the last time snapshot.  This
problem uses a base resolution of 60 zones, with eight allowed levels
of mesh refinement.
\label{fig:boostcol1}}
\end{figure}

\begin{figure}
\includegraphics[width=4.5in, angle=-90]{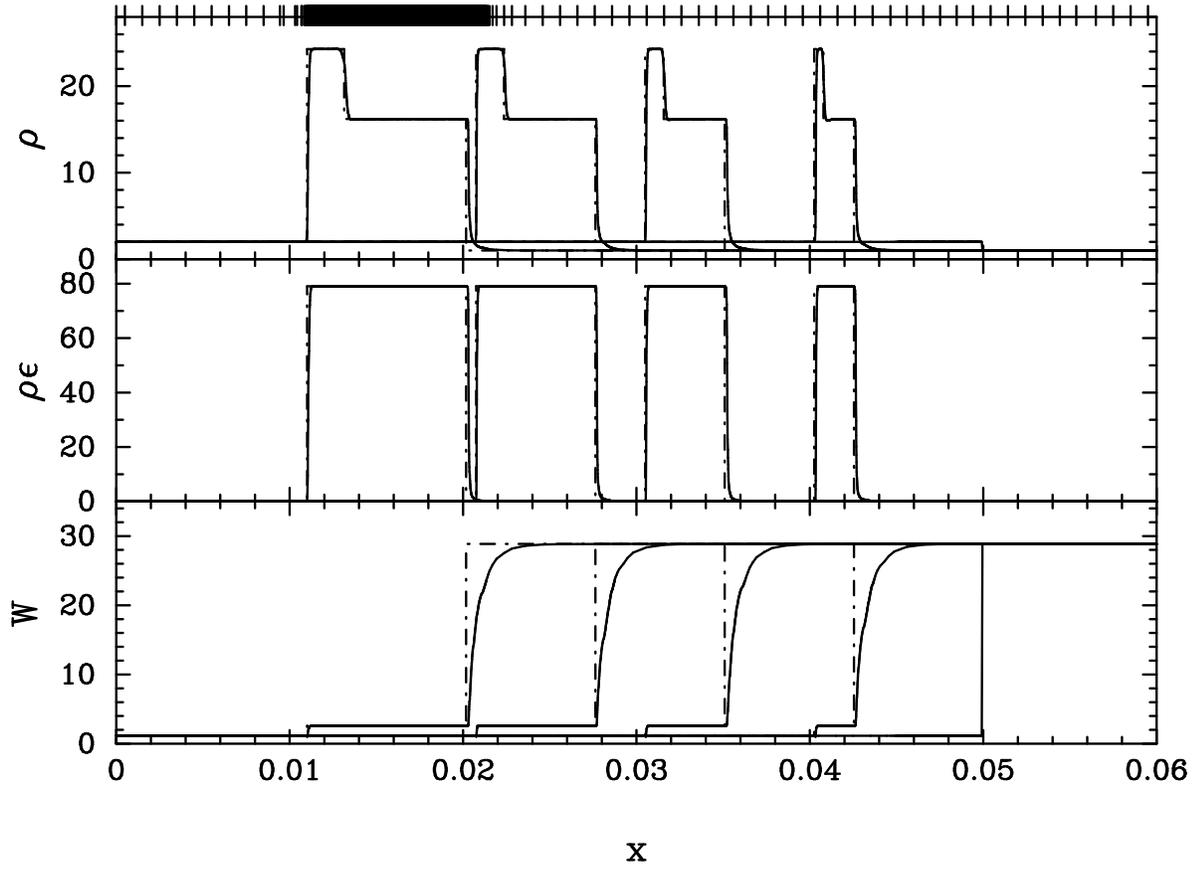}
\caption{As in Figure \ref{fig:boostcol1} except the two ideal fluids have
  unmatched initial proper densities $\rho_1 = 2$, $\rho_2 = 1$.
\label{fig:boostcol2}}
\end{figure}

\begin{figure}
\includegraphics[width=4.5in, angle=-90]{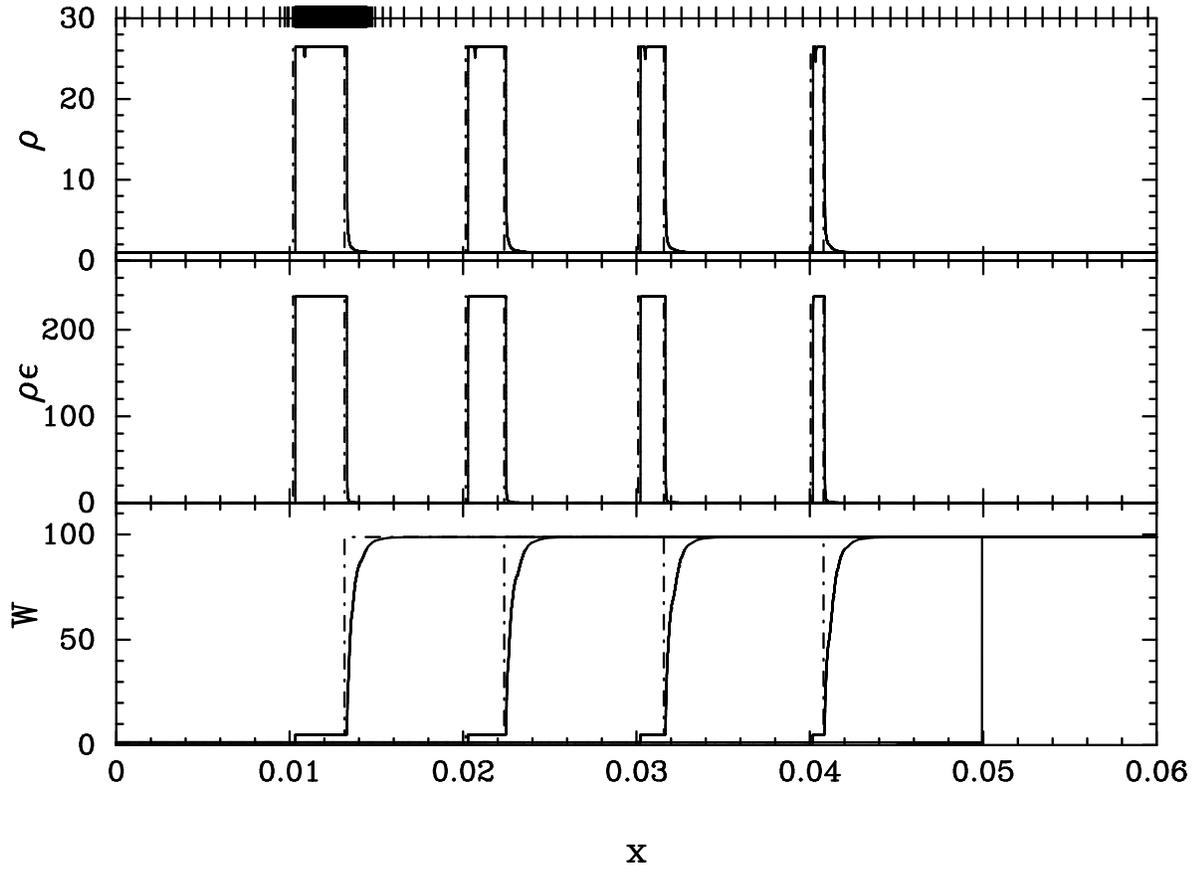}
\caption{As in Figure \ref{fig:boostcol1} except here $W' = 10$ and
the observer is boosted to the right at $W = 5$.  The base resolution
is 60 zones and 12 levels of mesh refinement are allowed.
\label{fig:boostcol3}}
\end{figure}

Figures \ref{fig:boostcol1}, \ref{fig:boostcol2} \&
\ref{fig:boostcol3} compare the analytic and numerical solutions,
using the eAV method, for three examples of boosted collisions.  We find very
good agreement to the solutions (Table \ref{tab:boostcol}) when the
opposing primed-frame momenta of each fluid is symmetric (Figure
\ref{fig:boostcol1}), asymmetric (Figure \ref{fig:boostcol2}) and for
a more extreme relativistic case with Lorentz factors of about 100
(Figure \ref{fig:boostcol3}).  Fractional errors for density
in the shocked region
are typically $\Delta \rho/\rho \sim 10^{-4}$ while both
proper energy and boost factor fractional errors are of order
$10^{-5}$.  The tail in mass density on the downwind (right hand) side
of the shocked region is a small proportion (3\% for Figures
\ref{fig:boostcol1} \& \ref{fig:boostcol2} and 1\% for Figure
\ref{fig:boostcol3}) of the total shocked mass and converges to zero
with increased zoning.  The NOCD method also gives good agreement, but
tends to be slightly more diffusive, with a larger mass tail on the trailing
shock.  In the most extreme relativistic case, shown in Figure
\ref{fig:boostcol3}, the numerical results are slightly behind (to the
right of) the analytical solution, corresponding to temporal delay in
the onset of the numerical shock of $\Delta t \approx 10^{-4}$. These
highly relativistic shocks require very fine zoning and thus AMR is
extensively used in order that these runs be computationally feasible.
Up to 12 levels of refinement are used in the highest boost test.
In these examples a simple density threshold is used, above which
refinement is triggered; however, more complex (and efficient)
triggering methods such as discussed in the next section have also
been successfully employed.

\subsubsection{Blast Waves}

A further application of the ultra-relativistic capabilities of the eAV and
NOCD methods demonstrated in the previous section and one that is of
particular astrophysical significance is the relativistic
blastwave.  Much like its Newtonian counterpart, the Sedov-Taylor
blastwave \cite[e.g. ][]{ll59}, there is a self-similar analytical
solution for the ultra-relativistic blastwave, first described by
\citet{bm76}.

The analytical solution of \citet{bm76} depends on the initial total
energy in the blastwave, $E_\text{BM}$, the Lorentz factor with which
its shock front initially expands, $\gamma_\text{BM}$, and the ambient
density, $\rho_\text{BM}$, into which it expands.  From these one can
define an initial radius of the blastwave
\begin{equation}
  r_\text{BM} \equiv \biggl(\frac{17 E_\text{BM}}{8\pi \gamma^2_\text{BM}
  \rho_\text{BM}} \biggr)^{1/3} ~.
\label{eqn:r_bm}
\end{equation}
The solution for radii, $0 < r < r_\text{BM}$, is based on the similarity
variable
\begin{equation}
  \chi(r) \equiv 1 + 8 \gamma^2_\text{BM} (1 - r/r_\text{BM}) ~.
\label{eqn:bm_chi}
\end{equation}
The coordinate density and energy, as used by \Cosmospp, and the Lorentz factor
are
\begin{eqnarray}
  D(r) &=& 2 \sqrt{-g} \rho_\text{BM} \gamma^2_\text{BM} \chi(r)^{-7/4} \\
  E(r) &=& \sqrt{-g} \rho_\text{BM} \gamma^2_\text{BM} (\sqrt{2} \gamma_\text{BM}
  \chi(r)^{-23/12} - 2 \chi(r)^{-7/4}) \\
  \gamma(r) &=& \sqrt{1 + \frac{\gamma^2_\text{BM}}{{2 \chi(r)}} }
\label{eqn:dega_bm}
\end{eqnarray}
where we have interpreted equation (29) of \citet{bm76} as the radial component of
the 4-velocity in order that $\gamma \ge 1$.  For all $r > r_\text{BM}$, $D(r)
= \sqrt{-g} \rho_\text{BM}$, $\gamma(r) = 1$ and the energy is set to
a numerically insignificant value, typically $E(r) = 10^{-4} D(r)$.

The relativistic blastwave is characterized by a very thin, $\Delta r
\propto r_\text{BM}/(8 \gamma^2_\text{BM})$ [eqn. \ref{eqn:bm_chi}],
shell of matter and energy rapidly expanding into an external medium.
Very fine zoning is required to resolve the shell, while relatively
few zones are necessary in the vacuous bubble enclosed by the shell or
the external medium.  Thus this problem is ideally suited for adaptive
mesh refinement.  For example, a blastwave with an initial Lorentz
factor of $\gamma_\text{BM} = 10$ would require 800 zones across its
radius just to ensure that the shell is represented by a single zone.
Because of the steep gradients behind the shock, to {\it resolve} the
shell requires $f_\text{res} \sim 100$ times as many zones, which
becomes a large computational problem.  This is compounded by the fact
that typical simulations evolve the blastwave over distances $\propto
r_\text{BM}$, thus requiring about $r_\text{BM}/\Delta r \sim
r_\text{BM} f_\text{res}/(k_{cfl} \Delta t) \sim 800
f_\text{res}/k_{cfl} \sim 10^6$ time steps, where $k_{cfl}$ is the
Courant factor.  For $\gamma_\text{BM} > 10$, stability typically
requires $k_{cfl} \sim 0.1$.  Therefore adaptive mesh refinement, with
zones concentrated into the shell, is a practical necessity for
relativistic blastwaves.

The relativistic blastwave, with its thin shell, steep gradients,
and sharp cusp at the shock front, is challenging to simulate.  The
method developed for these problems is to refine a zone if any of three
criteria are met:  First, all zones for which the density $\rho >
f_\text{thresh} \rho_\text{max}$ are refined to the fullest allowable
extent, where typically $f_\text{thresh} = 0.9$.  This condition is
required to keep the cusp at the shock front as resolved as possible.
Second, the normalized derivative of the proper density field,
$(\Delta x/\overline{\rho}) \nabla \rho$, must remain below a
threshold, where $\Delta x$ is the zone size and $\overline{\rho}$ is
the average value.  This is effective at maintaining the steep
gradient behind the shock.  Finally, the normalized curvature of this
field, $\Delta x |\nabla^2 \rho | / |\overline{\nabla \rho} |$, must
remain below a maximum threshold.  This condition puts zones ahead of
the advancing shock due to the high curvature at the shock
discontinuity, and also smooths the regions tagged by slope.  If $\rho
< f_\text{thresh} \rho_\text{max}$ and both normalized slope and
curvature of a zone fall below their respective thresholds,
then that zone is tagged for derefinement.

\begin{figure}
  \includegraphics[width=4.5in, angle=-90]{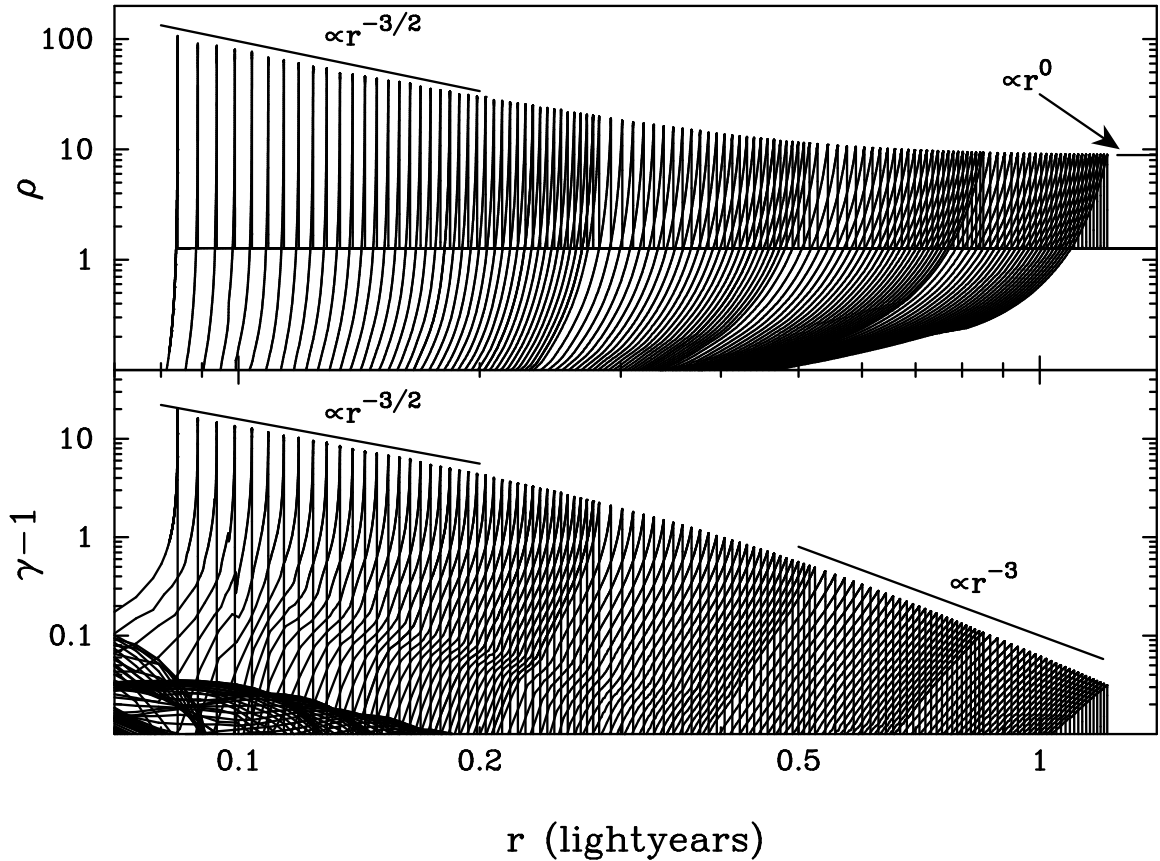} \caption{A
  spherical relativistic blastwave with energy $E_\text{BM} = 1 \text{
  foe} \equiv 10^{51}$ ergs, $\gamma_\text{BM} = 30$, plows into an
  external density of $\rho_\text{BM} = 1$ baryon cm$^{-3} = 1.27 $
  foe lightyear$^{-3}$ from an initial radius $r_\text{BM} = 0.084$
  lightyear.  The material immediately behind
  the shock is initially shocked to $\rho = 2^{3/2} \gamma_\text{BM}
  \rho_\text{BM}$, $\gamma = \gamma_\text{BM}/\sqrt{2}$
  and evolves as $\rho \propto \gamma - 1
  \propto r^{-3/2}$.  Once the blastwave decelerates to $\gamma \sim
  1$ it transitions to the non-relativistic Sedov-Taylor blastwave for
  which the shocked material evolves as $\rho = \rho_\text{BM}
  (\Gamma+1)/(\Gamma-1) = 8.9 $ foe lightyear$^{-3}$, $\Gamma = 4/3$,
  and $\gamma-1 \propto v^2 \propto r^{-3}$ \citep{ll59}.  This
  simulation uses the NOCD method.  For this figure
  the dump frequency was halved periodically to allow late time
  profiles to be resolved.  \label{fig:blastwave1}}
\end{figure}

As shown in Figure \ref{fig:blastwave1}, \Cosmospp is able to evolve a
blastwave with $E_\text{BM} = 10^{51}$ ergs from an initial Lorentz
factor of $\gamma_\text{BM} = 30$ to a non-relativistic blastwave.
Such a simulation is demanding, requiring $\sim 10^6$ cycles to evolve
the relativistic phase. This simulation is evolved on a base mesh of 100
zones with initially 17 levels of allowed refinement.  The number
of allowed refinement levels is periodically decremented nine times as the
blastwave decelerates and broadens. Thus only eight levels of
refinement are employed during the non-relativistic phase.  As a result,
the time step during the final non-relativistic phase is
$2^{10} \approx 1000$ times larger than the initial timestep.

\begin{figure}
  \includegraphics[width=4.5in, angle=-90]{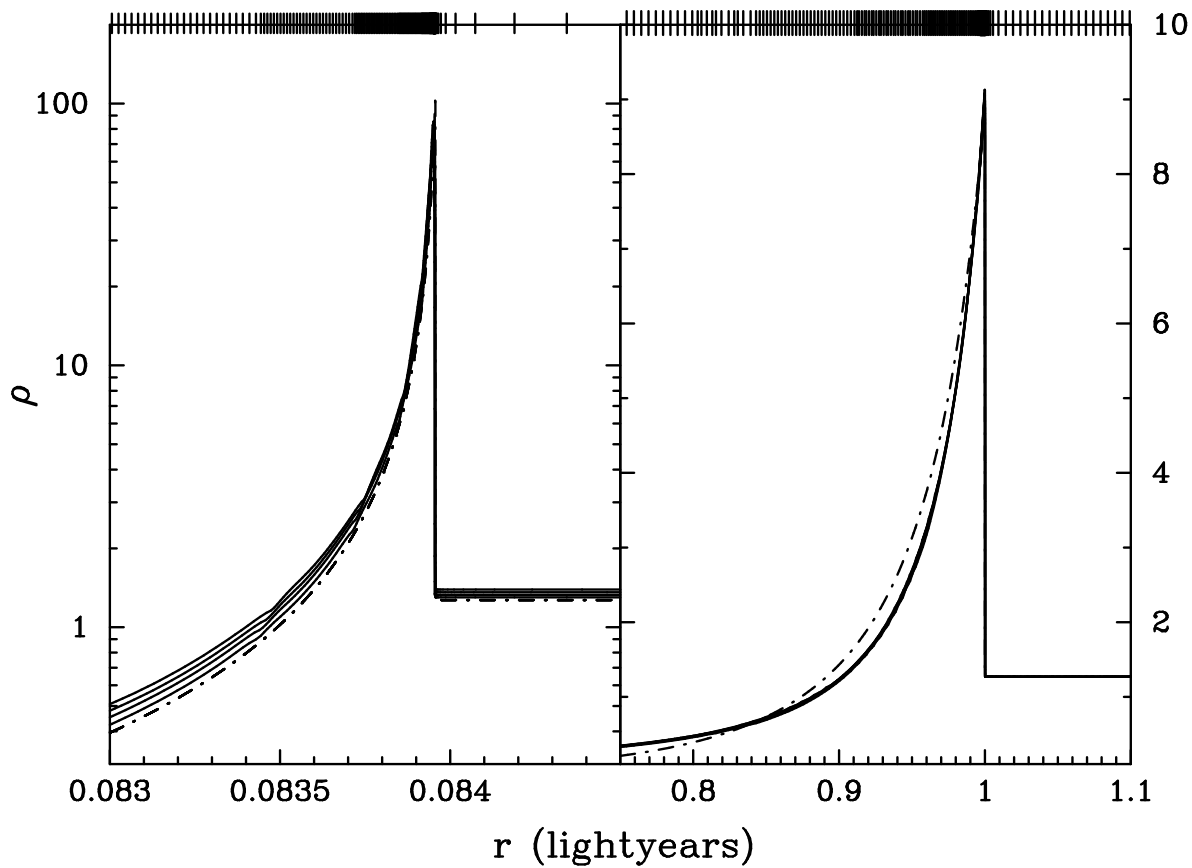}
  \caption{Self-similarity of the relativistic (left) and
  non-relativistic (right) phases of the blastwave of Figure
  \ref{fig:blastwave1}.  As discussed in the text, the relativistic
  (left) panel demonstrates the eAV method, while the non-relativistic
  (right) panel uses the NOCD method.  At left, the initial
  relativistic proper density profile, $\rho = D/W$, (dot-dashed line)
  of eqn.~(\ref{eqn:dega_bm}) is plotted with four subsequent profiles
  at $\sim 100,000$ cycle intervals.  Each profile is scaled in radius
  to align with the initial shock front and the density is scaled by
  $\rho \propto r^{-3/2}$.  Save an initial decrease in the magnitude
  of the peak of about 10\%, and a gradual increase in the density of
  the lower velocity tail, the self-similar profile is well
  maintained.  At right, the Sedov solution \cite[dot-dashed line,
  e.g.][]{ll59} with $\Gamma = 4/3$ is plotted with six density
  profiles from the non-relativistic phase of the run from Figure
  \ref{fig:blastwave1}.  The radii, ranging from 1 to 1.2 lightyears
  for times 1.33 to 2 years respectively, are scaled to the radius of
  one lightyear.  The shocked density $\rho = \rho_\text{BM}
  (\Gamma+1)/(\Gamma-1) = 8.9 $ foe lightyear$^{-3}$ is not scaled,
  but is naturally captured by the code to within a few percent.  The
  scaled profiles are virtually identical to each other and well
  reproduce the Sedov profile.  Tick marks at the top of the plots
  show the mesh nodes for a typical profile.  The left (right) plot
  has 17 (8) allowed levels of refinement.
  \label{fig:blastwave2}}
\end{figure}

Figure \ref{fig:blastwave2} shows the self-similarity of the
relativistic and non-relativistic phases of the blastwave evolution by
scaling and super-imposing several density profiles for each phase.
One can see excellent agreement with the analytical solution in both
cases.  For the relativistic phase, the eAV method (shown in the left
plot) tends to more robustly evolve the analytical solution, with less
sensitivity to time step or AMR refinement prescriptions.  However, the
NOCD method (shown in the right plot) more reliably transitions to the
non-relativistic limit; for instance capturing the shock discontinuity
in density with $\sim 1$ \% accuracy as opposed to $\sim 10$ \%
accuracy for the eAV method.  This might be due to the enforced energy
conservation of the NOCD method.  The eAV method typically only loses
1 -- 2\% of the total energy over the course of the evolution, but
this may occur primarily in the shocked shell, thus causing a slight
drift in the solution.


\subsection{Magnetohydrodynamics}
\label{sec:mhdtests}

\subsubsection{Alfv\'en Wave Propagation}
\label{subsubsec:alfven}

The class of linear Alfv\'en waves described by \cite{dev03a}
provide an excellent test of the method of characteristics for
magnetic fields subject to transverse or shearing mode perturbations.
Considering a general wave function $f(x-v_A^\pm t)$,
solutions to the linear perturbation equation with a fixed
background field ${\cal B}^x$ and constant velocity $V^x$
in Minkowski spacetime yields for the transverse components
\begin{eqnarray}
V^y (x,t)        &=& \left(\frac{1-\zeta\chi}{2}\right) f(x-v_A^- t)
                   + \left(\frac{1+\zeta\chi}{2}\right) f(x-v_A^+ t) ~, \\
{\cal B}^y (x,t) &=& \frac{\zeta}{2}\left[ f(x-v_A^- t) - f(x-v_A^+ t)\right] ~,
\end{eqnarray}
with parameters
\begin{equation}
\zeta  =  \frac{{\cal B}^x (1+\eta^2)}{\eta\sqrt{\eta^2+W^{-2}}} ~, \quad
\eta^2 =  \frac{||B||^2}{4\pi \rho h W^2} ~, \quad
\chi   = -\frac{\eta^2 V^x}{{\cal B}^x(1+\eta^2)} ~, \quad
\beta  =  \frac{8\pi P}{||B||^2}
~,
\end{equation}
and Alfv\'en speed
\begin{equation}
v_A^\pm = \frac{V^x \pm \eta \sqrt{\eta^2+W^{-2}}}{1+\eta^2}
        ~\stackrel{\displaystyle\longrightarrow}{{\scriptstyle V^x = 0}}~
        \sqrt{\frac{||B||^2}{4\pi \rho h + ||B||^2}}
        = \sqrt{\frac{2(\Gamma-1)\epsilon}
               {\beta(1+\Gamma\epsilon) + 2(\Gamma-1)\epsilon}}
~.
\end{equation}
We consider two cases of linear Alfv\'en waves: a stationary
background ($V^x=0$, case A) in which the pulses travel in opposite
directions with equal amplitudes, and a moving background ($V^x =
0.1c$, case B) where the pulses split into asymmetrical waves. These
two cases correspond to Models ALF1 and ALF3 of \citet{dev03a}. In
both cases, the fluid is initialized with a uniform density
$\rho=1$, specific energy $\epsilon = 10^{-2}$, and adiabatic index
$\Gamma=5/3$.  The transverse magnetic field components (${\cal B}^y
= {\cal B}^z$) are initially zero. The longitudinal magnetic field
component (${\cal B}^x$) is set by our choice of $\beta$
($\beta=0.001$ for case A and 0.01 for case B). The transverse
velocity is initialized with a square pulse: $V^y = 10^{-3}$ for $1
< x < 1.5$, $V^y = -10^{-3}$ for $1.5 \le x < 2$, and zero
elsewhere. The problem is run on a grid 3 units in length.

\begin{figure}[htb]
\plottwo{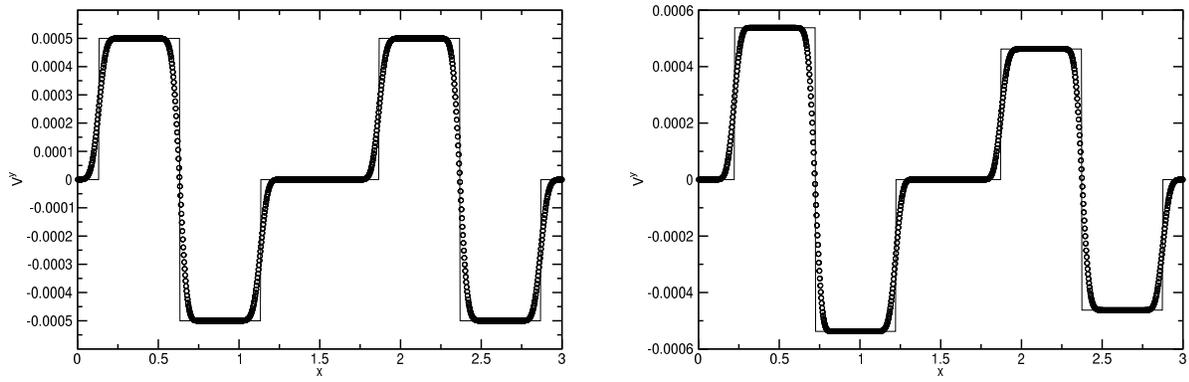}{f11b.eps}
\caption{Tranverse velocity $V^y$ for two cases of the
Alfv\'en wave test: ({\em a}) case A at $t=0.9$ and ({\em b}) case B at
$t=1.1$. The numerical resolution is 1024 zones;
the solid lines are the analytic solutions.
}
\label{fig:alf_vel}
\end{figure}

For case A, the Alfv\'en velocities are $\vert v_A^\pm \vert =
0.963$ ($W=3.76$); for case B, $v_A^+=0.792$ ($W^+=1.64$) and
$v_A^-=-0.705$ ($W^-=1.41$). Figure \ref{fig:alf_vel} shows the
numerical results for the artificial viscosity (AV) method on a grid with
1024 zones overlaid with the analytic solution. The eAV method gives
identical results for this test and is not shown. The errors in the
plateaus for the stationary background case (case A) are extremely
small ($<0.005$\%); for the moving background (case B), the errors
are somewhat larger, though still quite small ($<0.1$\%)

\subsubsection{Magnetosonic Shock Tube}

\begin{figure}[htb]
\plotone{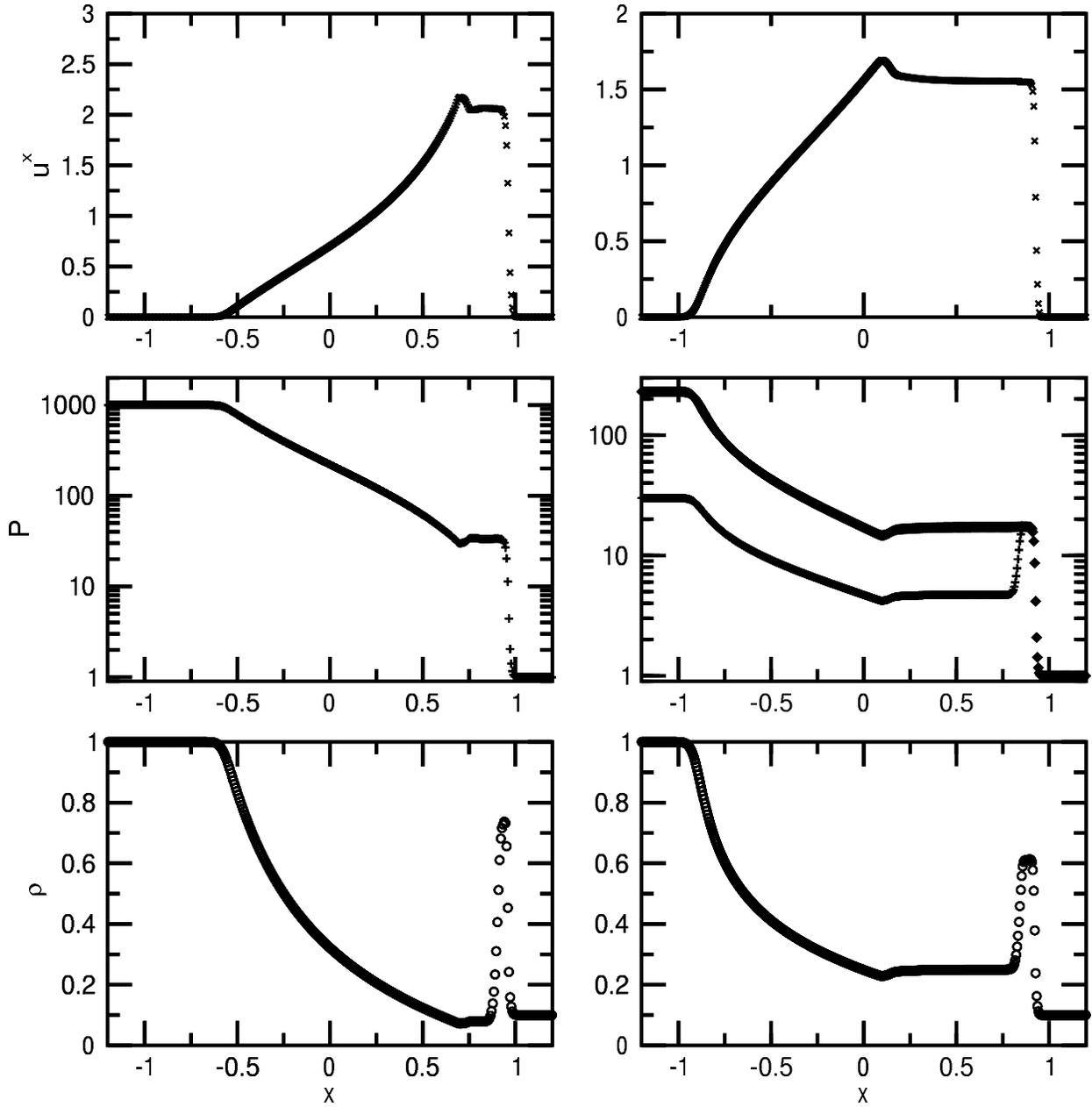}
\caption{{\em Left panel:} Magnetosonic shock tube problem 1 from \citet{kom99}.  The
magnetic field is normal to the initial discontinuity (at $x=0$) so it
does not play a dynamical role.  The symbols show the numerical solution
at $t=1$ for 400 zones using the AV method.
{\em Right panel:} Magnetosonic shock tube problem 2 from \citet{kom99}.  The
initial magnetic field to the left of $x=0$ is parallel to the discontinuity,
while the field is absent to the right.
The symbols show the numerical solution
at $t=1$ for 500 zones using the AV method.}
\label{fig:mhdtube_AV}
\end{figure}

\begin{figure}[htb]
\plotone{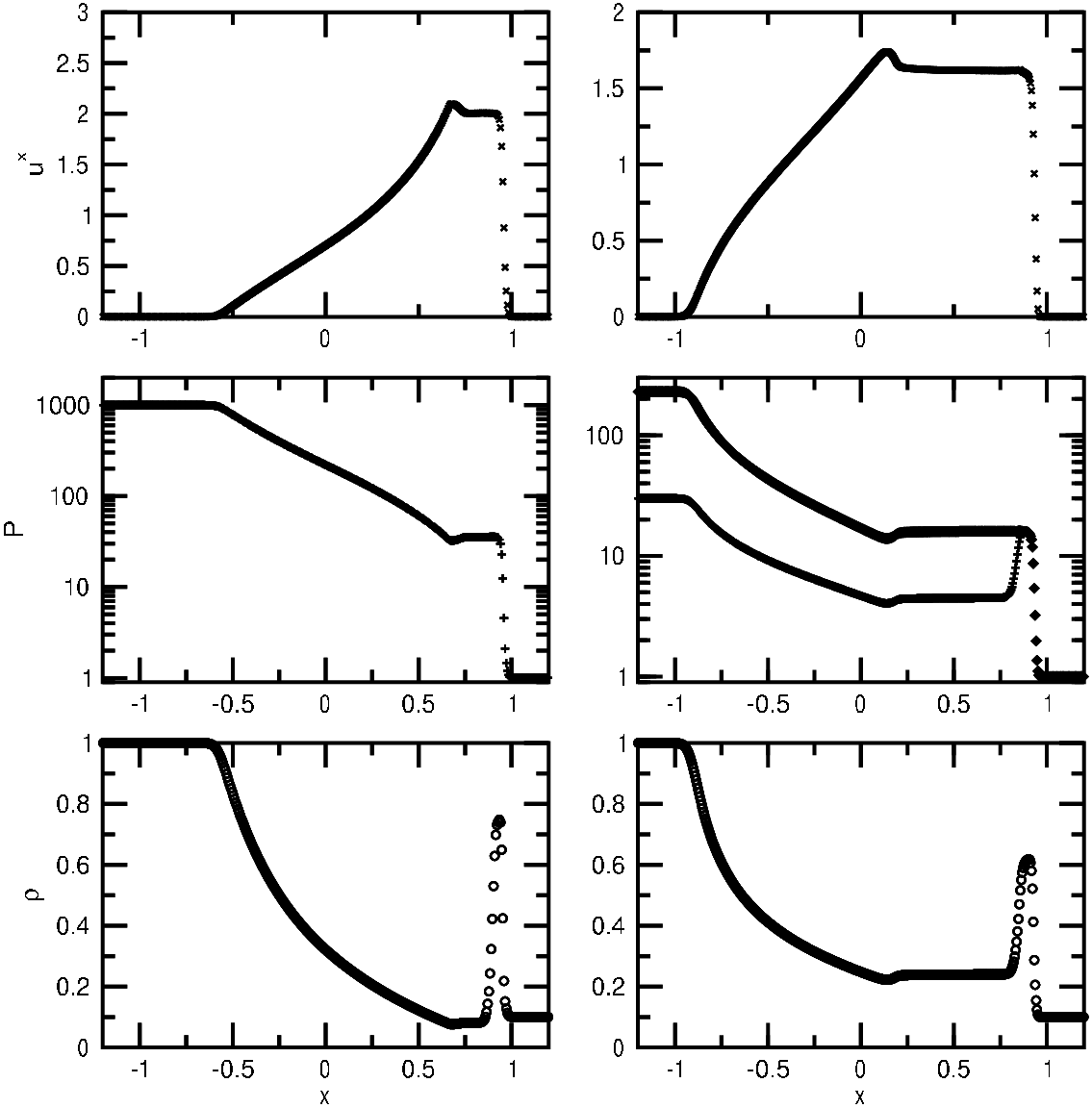}
\caption{As Figure \protect{\ref{fig:mhdtube_AV}} but with
the extended viscosity (eAV) scheme.}
\label{fig:mhdtube_eAV}
\end{figure}

\begin{figure}[htb]
\plotone{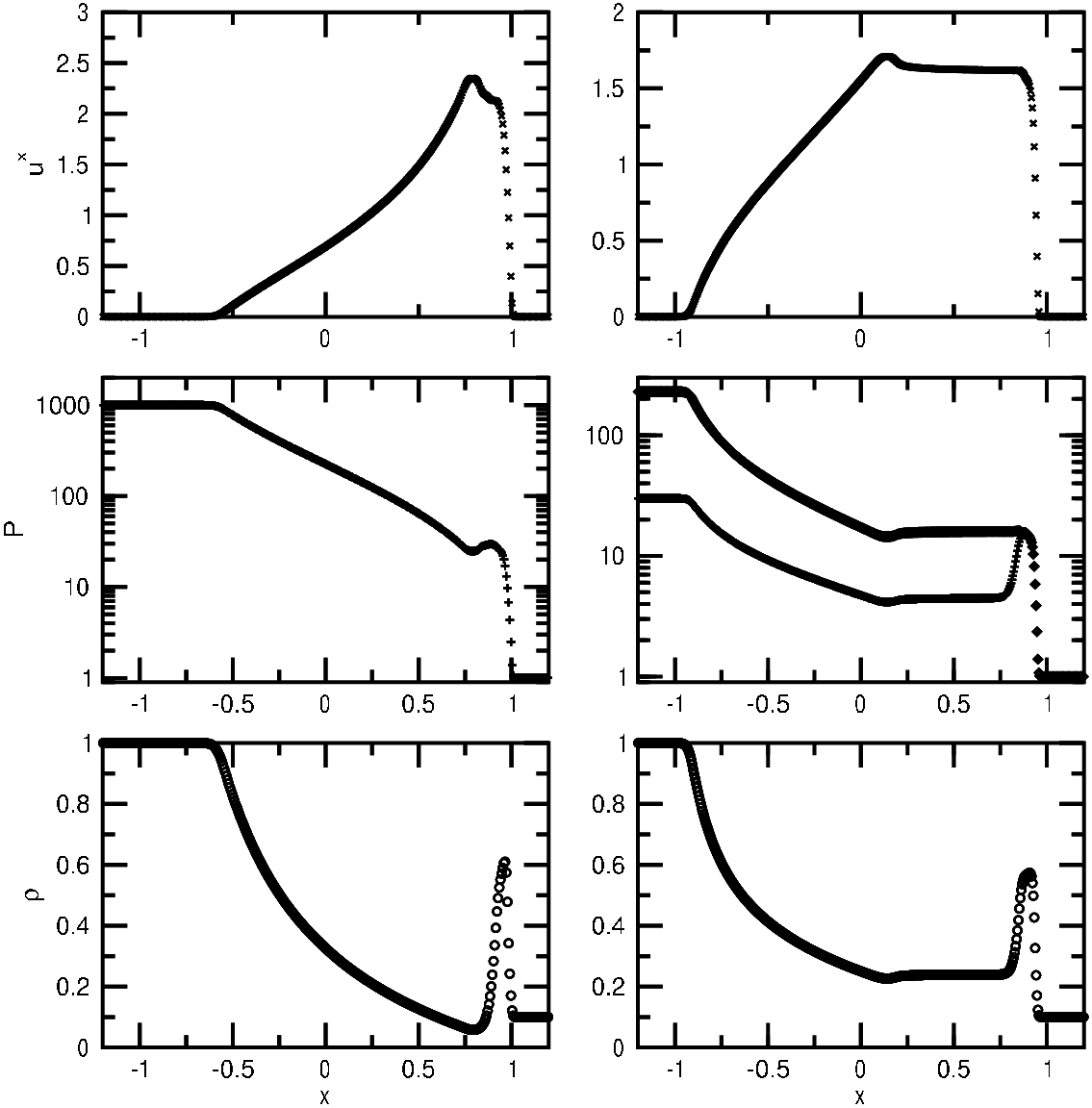}
\caption{As Figure \protect{\ref{fig:mhdtube_AV}} but with
the non-oscillatory central difference (NOCD) scheme.}
\label{fig:mhdtube_NOCD}
\end{figure}

Next we perform another set of numerical shock tube tests, this time
including magnetic fields. Similar to the hydrodynamic shock tube
test, these problems combine strong shocks and rarefaction features,
thus fully stressing the code in the limit of flat space. We
initialize two versions of this test, both from \citet{kom99}. In
the first test, the initial state is specified as $\rho_L = 1$, $P_L
= 10^3$, $V_L = 0$, ${\cal B}^x_L=1$, ${\cal B}^y_L = {\cal B}^z_L =
0$ to the left of the membrane and $\rho_R = 0.1$, $P_R = 1$, $V_R =
0$, ${\cal B}^x_R=1$, ${\cal B}^y_R = {\cal B}^z_R = 0$ to the
right.  Note that since the magnetic field in this case is parallel
to the flow and continuous across the discontinuity, it should not
play a dynamical role.  In this sense, the problem is really a
hydrodynamical shock tube similar to the one considered in \S
\ref{sec:stube}.  Nevertheless, this test appears frequently in the
literature and is worth reproducing here to confirm consistency in
the MHD solver. For the second test, the initial state is $\rho_L =
1$, $P_L = 30$, $V_L = 0$, ${\cal B}^y_L=20$, ${\cal B}^x_L = {\cal
B}^z_L = 0$ to the left of the membrane and $\rho_R = 0.1$, $P_R = 1$,
$V_R = 0$, ${\cal B}^x_R = {\cal B}^y_R = {\cal B}^z_R = 0$ to the
right.  Here, the initial discontinuity of the field allows it to
play a dynamical role in the rarefaction and contact wave regions.
These tests are run using the scalar artificial viscosity with a
quadratic viscosity coefficient $k_q=2.0$, Courant factor
$k_{cfl}=0.3$, and are evolved on fixed, uniform grids. Case 1 uses
a linear viscosity coefficient $k_l=0.3$, whereas case 2 does not
use the linear term ($k_l=0$). We do not calculate the analytic
solutions for these tests, although a direct comparison can be made
between our Figures \ref{fig:mhdtube_AV}, \ref{fig:mhdtube_eAV}, and
\ref{fig:mhdtube_NOCD}, and Figure 6 of \citet{kom99} since we use
the same resolution (400 and 500 zones for shock tubes 1 and 2,
respectively) as that work. We point out that our different
methodologies all give similar results as the Godunov-type scheme
used in \citet{kom99}, showing many of the same pathologies (the
small post-shock features in shock tube 1 and the kink at the
rarefaction edge in shock tube 2), although our AV and eAV methods
appear to do better at capturing the density plateau in the
Lorentz-contracted shell of material immediately behind the shock.

\subsubsection{Bondi Flow}

As a test of hydrodynamic and MHD flows in spacetimes with
nontrivial curvature, we first consider radial accretion onto a
compact, strongly gravitating object, in this case a Schwarzschild
black hole.  The analytic solution \citep{mic72} is characterized by
a critical point in the flow as
\begin{eqnarray}
(u^r_c)^2 & = & M/2r_c ~, \\
V_c^2 & = & (u^r_c)^2/[1-3(u^r_c)^2] =
\frac{(1+n)T_c}{n[1+(1+n)T_c]} ~,
\end{eqnarray}
where $u^r$ is the radial component of the fluid 4-velocity, $V_c$
is the sound speed at the critical point, $M$ is the mass of the
black hole, $n=1/(\Gamma - 1)$ is the polytropic index, and
$T=P/\rho$. The remainder of the flow is described through the
equations
\begin{eqnarray}
T^n u^r r^2 & = & C_1 ~, \\
\left[ 1+(1+n)T \right]^2 \left[ 1- \frac{2M}{r} + (u^r)^2 \right] &
= & C_2 ~.
\end{eqnarray}
Following \citet{haw84a}, we fix $C_1$ and $C_2$ by choosing the
critical radius of the solution $r_c=8GM/c^2$ and fixing
$\Gamma=4/3$. We also fix the density at the critical radius ($\rho_c$)
such that $\dot{M}=4\pi r_c^2 \rho_c u^r_c = -1$.

The physical domain of our simulation extends from $r=0.98r_{BH}$ to
$r=20GM/c^2$, where $r_{BH}=2GM/c^2$ is the radius of the black-hole
horizon. Here we use Kerr-Schild coordinates, which allow us to
place the inner boundary of the computational domain inside the
horizon. The radial coordinate is replaced by a logarithmic
coordinate $\eta=1+\ln(r/r_{BH})$, and the problem is evolved over a
time interval $\Delta t = 100 GM/c^3$. We measure the convergence of
our solution at four different resolutions (32, 64, 128, and 256
zones) using the one-dimensional $L$-1 norm of $\rho$. For 256
zones, the respective errors for the AV, eAV, and NOCD methods are
$1.14\times 10^{-3}$, $1.08\times 10^{-3}$, and $4.66\times
10^{-4}$. The convergence rates are between first and second order:
1.3, 1.4, and 1.4 for the AV, eAV, and NOCD methods, respectively.
These rates reflect a slight degradation of truncation order as
expected for curvilinear meshes since we do not currently construct
volume-centered centroids nor high order face-centered interpolants
that affect gradient operators, for example. Nevertheless, these
convergence rates are generally consistent with the convergence
reported by \citet{dev03a}.

We can extend this test to the GRMHD regime by adding a radial
magnetic field. Inclusion of such a field (satisfying $\partial_r
\mathcal{B}^r = 0$) does not alter the analytic solution for any of
the primitive fields ($\rho$, $P$, or $u^r$). We should point out,
however, that this treatment does not satisfy the whole set of
Maxwell equations \citep{anton05}, yet it serves as a non-trivial
numerical test of the magnetic field terms in the code and is well
documented in the literature. The magnitude of the magnetic field is
set by $\vert\vert B^2 \vert\vert / (4\pi \rho) = 10.56$ ($\beta
\approx 1$) at $r=r_c$. With magnetic fields included, the
magnitudes of the errors increase to $3.89\times 10^{-3}$ and
$1.36\times 10^{-3}$ for the AV and eAV methods with 256 zones, but
the convergence rates remain the same - 1.3 and 1.4, respectively.
We note that the divergence error in the magnetic field does not
build up appreciably during the course of these runs - $\sum
|\partial_i \mathcal{B}^i | \lesssim 10^{-14}$ in all cases. Since
we intend primarily to use either the AV or eAV methods for this
type of research application, we have not extended the NOCD
algorithm to account for the proper characteristic
magnetohydrodynamic speed in arbitrarily curved spacetimes, so we do
not report results from that method in this or the following
section.



\subsubsection{Magnetized Black Hole Torus}

\begin{figure}[htb]
\begin{tabular}{ccc}
\includegraphics[width=2 in]{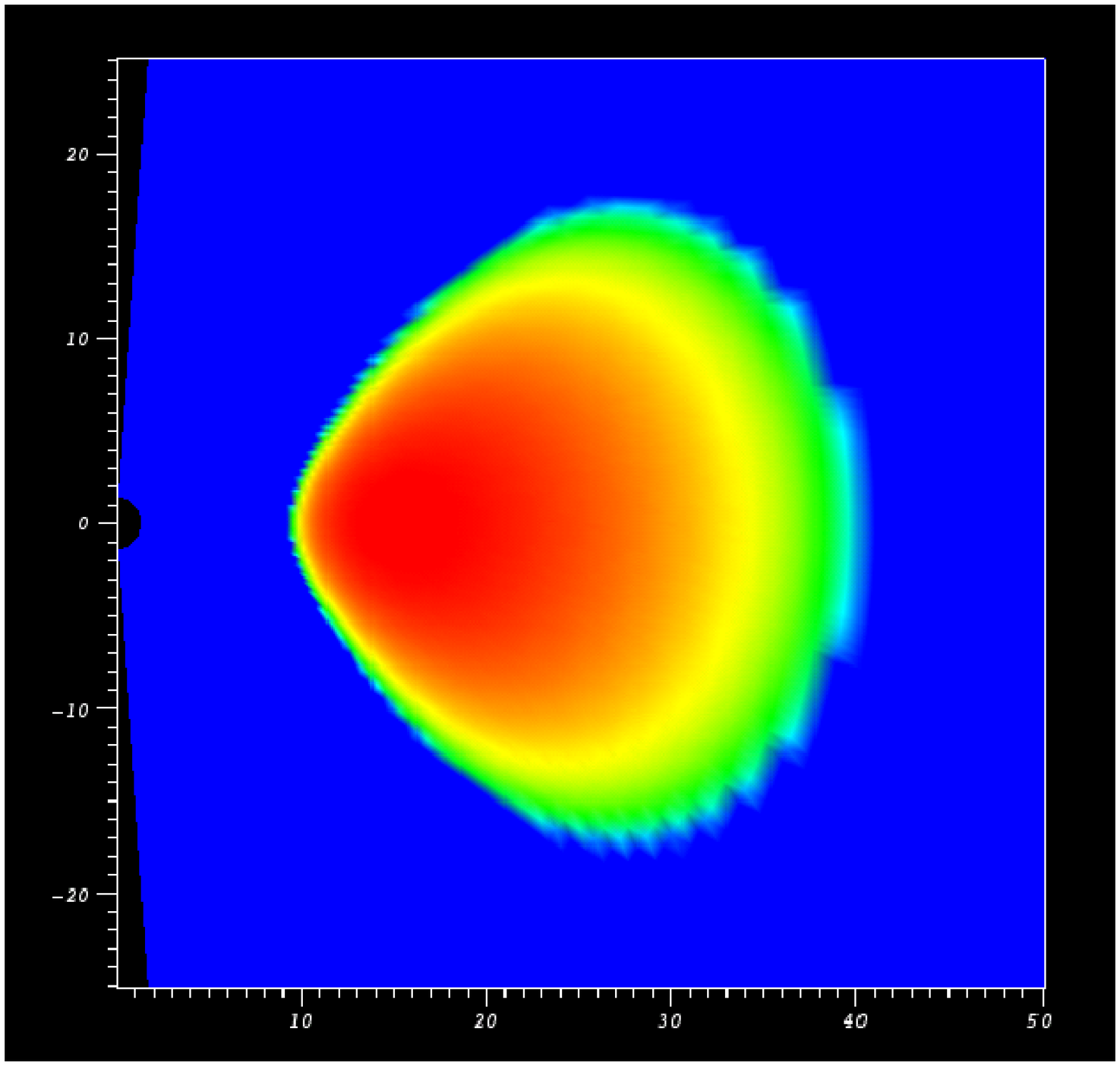} & \includegraphics[width=2 in]{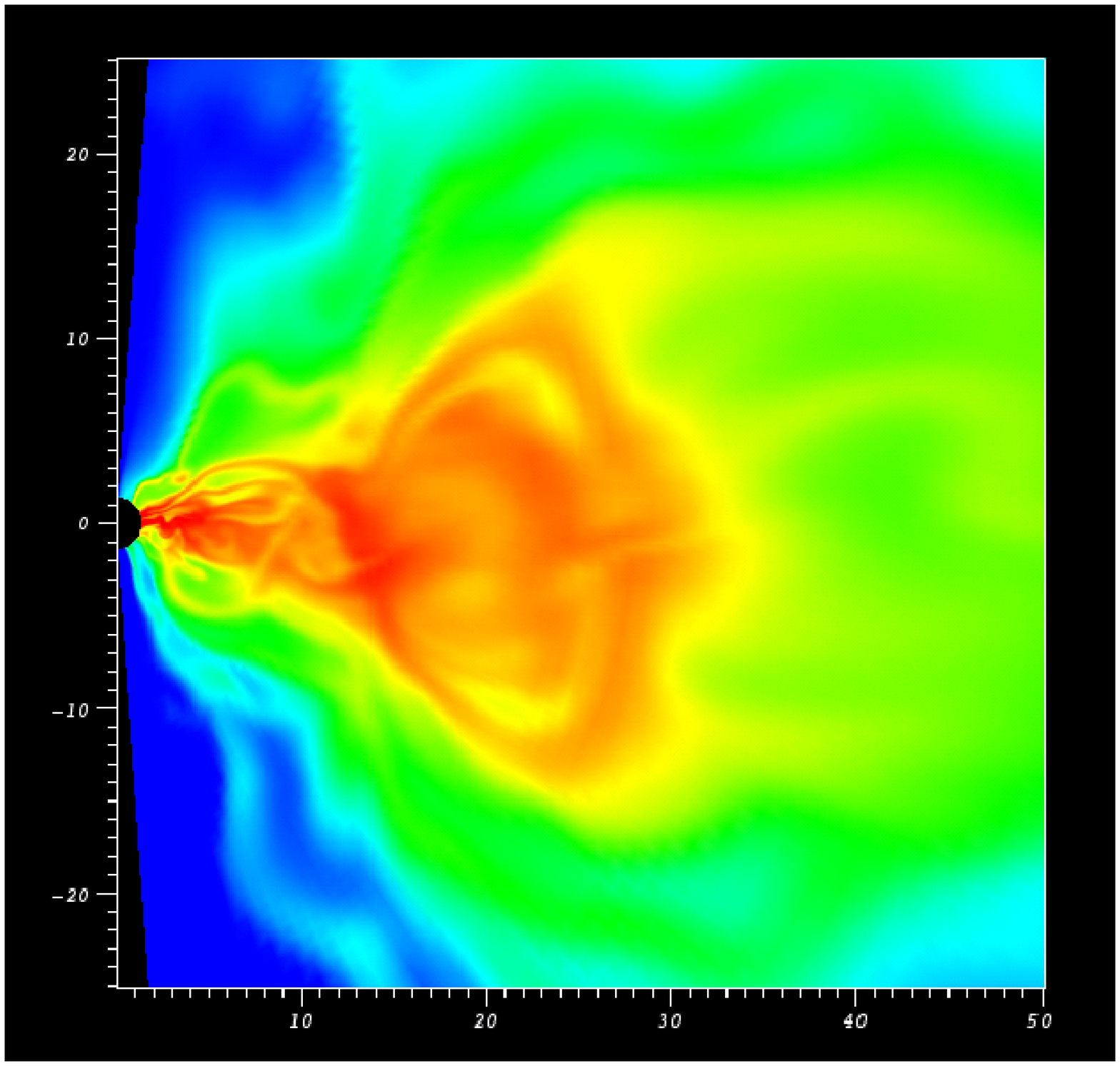} & \includegraphics[width=2 in]{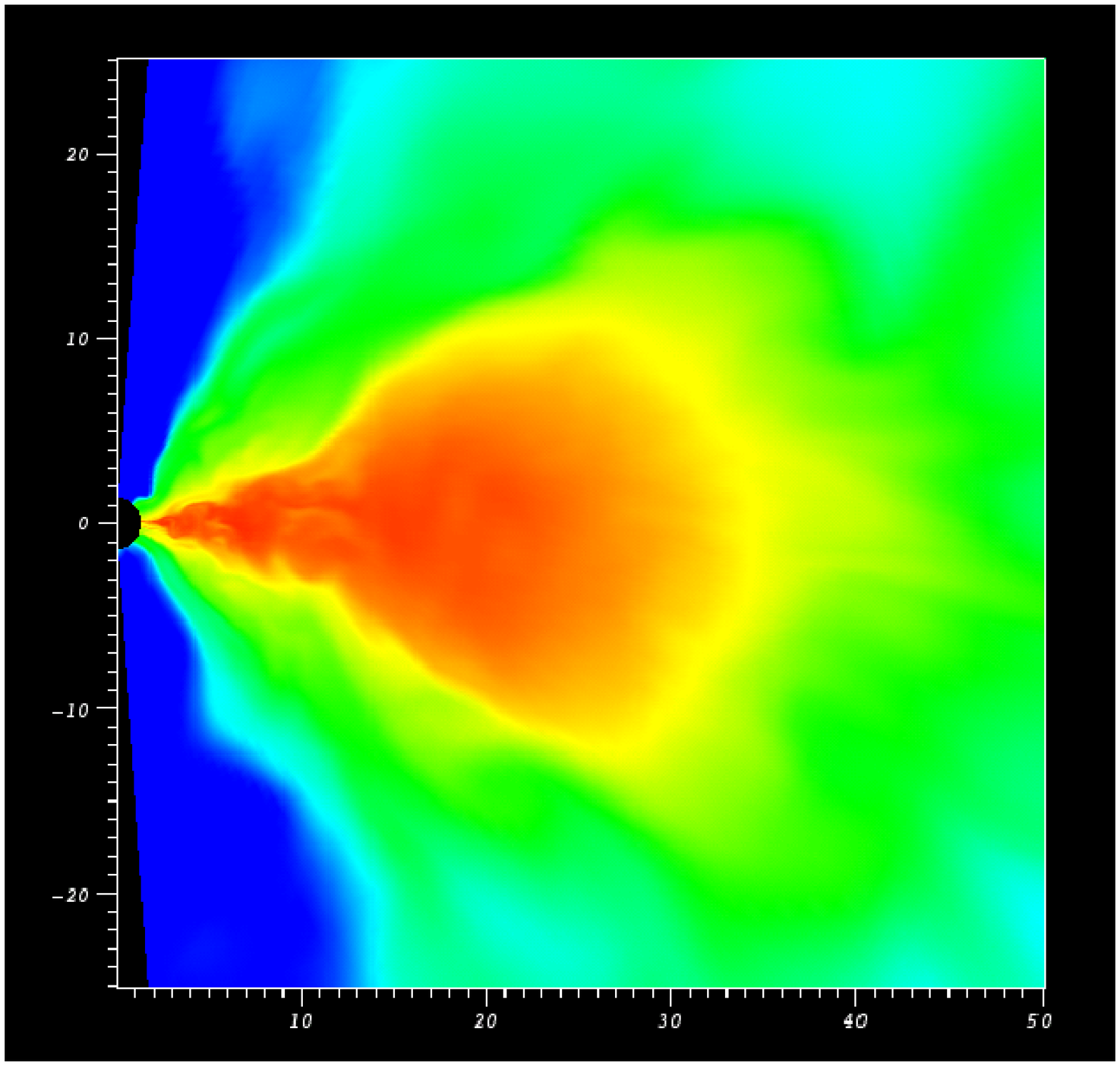}
\end{tabular}
\caption{Plot of gas density $\rho$ at ({\em a}) $t=0$, ({\em b})
$t=1250 M=3 \tau_{orb}$, and ({\em c}) $t=2300 M=6 \tau_{orb}$ for
the Kerr-Schild form of the metric. The density is scaled
logarithmically over 4 orders of magnitude and maintains the same
scale in all three panels.} \label{fig:mhd_torus_KS}
\end{figure}

Next we consider the astrophysically interesting problem of a
magnetized torus of gas orbiting a rotating black hole.  Although
there is no known analytic solution for this problem, it is
sufficiently well documented in the literature
\citep[e.g.][]{dev03a,gam03a,anton05} to serve as a useful test of the
code. It also represents one class of problems for which \Cosmospp
is intended to be used. We set this problem up using the same
parameters as model SFP-2D of \citet{dev03b}; specifically, the spin
of the black hole is $a/M=0.9$, the specific angular momentum of the
torus is $l/M=4.3$, the surface potential of the torus is
$(u_t)_{in} = -0.98$, and the equation of state of the gas is set by
the polytropic constant $\kappa=0.01$ and index $\Gamma=5/3$. The
initial setup of the torus is illustrated in Figure
\ref{fig:mhd_torus_KS}({\em a}), which shows a plot of the logarithm
of $\rho$.

To this torus we add initially weak poloidal magnetic field loops to
seed the magneto-rotational instability \citep[MRI, ][]{bal91}. The
initial magnetic field vector potential is \citep{dev03a}
\begin{equation}
A_\phi = \left\{ \begin{array}{ccc}
          k(\rho-\rho_{cut}) & \mathrm{for} & \rho\ge\rho_{cut}~, \\
          0                  & \mathrm{for} & \rho<\rho_{cut}~.
         \end{array} \right.
\label{eq:torusb}
\end{equation}
The non-zero spatial magnetic field components are then
$\mathcal{B}^r = - \partial_\theta A_\phi$ and $\mathcal{B}^\theta =
\partial_r A_\phi$.  These poloidal field loops coincide with the
isodensity contours of the torus. The parameter
$\rho_{cut}=0.5\rho_{max}$ is used to keep the field a suitable
distance inside the surface of the torus. Using the constant $k$ in
equation (\ref{eq:torusb}), the field is normalized such that
initially $\beta=P/(\vert \vert B \vert \vert^2/8\pi) \ge \beta_0=2$
throughout the torus. This initialization is slightly different than
\citet{dev03b}, who use a volume integrated $\beta$ to set the field
strength; the difference is such that $\beta_0=100$ in their work is
roughly comparable to $\beta_0=2$ here.


In the background region not specified by the torus solution, we
initially set up a cold ($e = 10^{-6}e_{max}$), low-density
($\rho = 10^{-6}\rho_{max}$), static ($V^i=0$) gas, where $e_{max}$
and $\rho_{max}$ are the initial internal energy and mass densities
at the pressure maximum of the torus (at $r_{center}=15.3 GM/c^2$). 
These values for $e$ and $\rho$
are then used as floors throughout the simulation. Obviously once
the evolution begins, the background is no longer static. At first
it falls toward the hole, but ultimately most of the background
region is filled with a magnetized "corona" and a low density, high
$\beta$ outflow launched by pressure forces and the opening of
magnetic field lines. Because the background gas always remains very
low density, it does not have a significant dynamical effect on the
torus, although its role in real astrophysical systems is still not
well understood.

For this test, we perform a two-dimensional, axisymmetric simulation
using our AV method. We restrict this test to a single level
fixed mesh with a resolution of $128^2$. In later applications, we
plan to employ the full three-dimensional, refined grid capabilities
of \Cosmospp. This simulation uses a logarithmic radial coordinate
of the form $\eta=1+\ln(r/r_{BH})$ and a concentrated latitude
coordinate $x_2$ of the form $\theta = x_2 + \frac{1}{2}(1-h)\sin(2
x_2)$ with $h=0.5$. The grid covers the angular scale $0.02\pi \le
\theta \le 0.98\pi$ and has radial boundaries at $r_{min}=0.98
r_{BH}$ and $r_{max}=120 M$. We use outflow boundaries at both
locations (radial fluid motion onto the grid is not allowed). The
chosen grid resolution gives a zone spacing of $\Delta r \approx
0.05 GM/c^2$ near the inner radial boundary and $\Delta r \approx
0.5 GM/c^2$ near the initial pressure maximum of the torus. This can
be compared with the characteristic wavelength of the MRI,
$\lambda_{MRI} \equiv 2\pi v_A/\Omega \approx 2.5 GM/c^2$ near the
initial pressure maximum.

\begin{figure}[htb]
\plotone{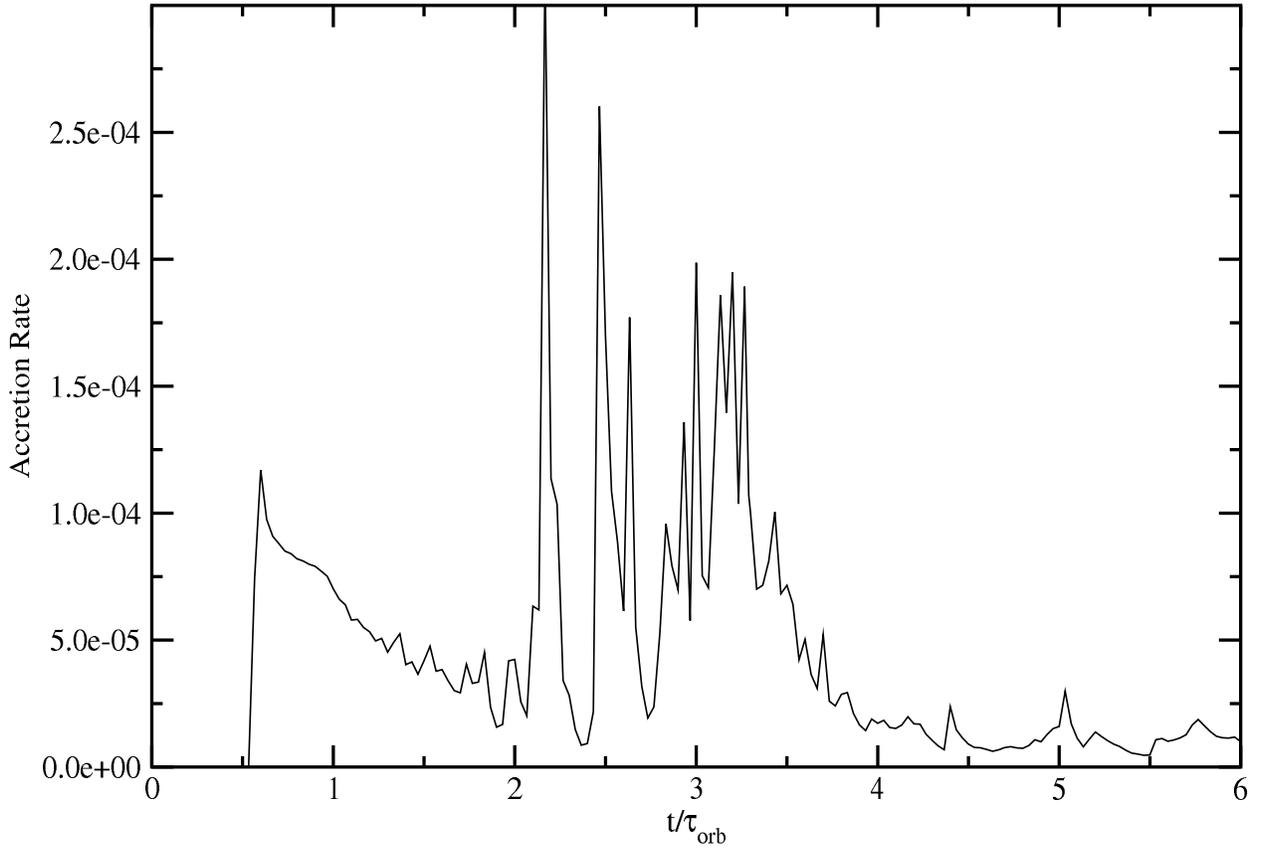} \caption{Mass accretion rate $\dot{M}$ at the
inner radial boundary $r_{min}$ normalized by the initial mass of
the torus.} \label{fig:mdot}
\end{figure}

Figure \ref{fig:mhd_torus_KS} shows the mass density distribution at
$t/(GM/c^3)=0$, 1250, and 2300 (roughly 0, 3, and 6 orbits at the
initial pressure maximum). For a more quantitative comparison,
Figure \ref{fig:mdot} shows the time history of the mass accretion
rate through the inner radial boundary of the grid. Comparing with
Figure 18 of \citet{dev03b}, we see similar amplitudes and
frequencies of variability in the accretion flow, particularly for
the first 4 orbits. We also find a similar value for the total mass
accretion: $\Delta M/M_0 = 0.13$ after 6 orbits here versus 0.14
after 10 orbits in \citet{dev03b}.
The accretion tapers off to a low value after about
3.5 orbits as expected since MRI
turbulence cannot be sustained indefinitely in an axisymmetric
simulation, due to Cowling's antidynamo theorem.
Throughout this simulation the divergence of the magnetic field
is held below a tolerable limit ($\sum |\partial_i \mathcal{B}^i | \sim 0.01$)
through the use of the divergence cleanser. Importantly, this
divergence error does not increase with time.

\section{Discussion}
\label{sec:conclusion}

We have developed a new object oriented code \Cosmospp for solving the fully
general relativistic magnetohydrodynamics equations on adaptive, unstructured meshes
using discrete finite volume and dimensionally unsplit methods.
Three basic numerical schemes have been
implemented and tested using both internal and total conservative energy formulations:
The first evolves internal energy with artificial viscosity for shock capturing;
the second uses a nonoscillatory central difference scheme to solve the
fully conservative (energy and momentum) form of equations; and the third
approach combines the internal and conservative energy equations with
artificial viscosity methods to achieve greater accuracy in highly
relativistic regimes.

We find by comparing the different methods presented here with other published
results, including Riemann-solver based codes, that, despite their simplicity,
artificial viscosity methods perform quite well for low to moderately
boosted flows, ($V/c \simeq 0.95$ in the shock tube and wall shock tests).
This is consistent with our conclusions from earlier work using dimensionally
split, finite difference methods on structured grids \citep{ann03a}.
However, it is well known that for higher velocity flows, traditional AV methods
tend to break down. We have demonstrated that the basic AV approach can be
extended easily to the ultra-relativistic regime by simply incorporating
a dual energy formalism (the eAV method) to solve both internal and conservative
energy equations with standard operator split procedures. This eAV procedure
results in significantly improved shock and wave capturing capabilities
that allows AV schemes to model flows at arbitrarily high boosts.
We have presented stable, accurate solutions for strong shock collision interactions
with boosts (velocities) easily exceeding $200$ ($0.99999 c$).

The NOCD method implemented here, although of a slightly different family
of algorithms than we adopted in our previous code, also
works quite well in the high boost regime. In fact this method has
significantly less numerical diffusion and is less sensitive to
Courant restrictions than our earlier implementation \citep{ann03a}.
The eAV and NOCD methods thus provide robust alternatives to simulating
highly relativistic flows since they are comparable in accuracy,
over the entire range of velocities we have simulated, to more
complicated Riemann-based codes. A similar conclusion regarding
central difference schemes was reached by \cite{ann03a} and \cite{ser04}.
Here we have extended the scope of tests and validity of
this class of methods to relativistic magnetohydrodynamics.

\begin{acknowledgements}
The authors would like to thank the VisIt development team at
Lawrence Livermore National Laboratory (http://www.llnl.gov/visit/),
especially Hank Childs, for visualization support.
This work was performed
under the auspices of the U.S. Department of Energy by
University of California, Lawrence
Livermore National Laboratory under Contract W-7405-Eng-48.
Funding support for P.C.F. was also provided by NSF grant AST 0307657.
Support was also provided by the National Science
Foundation under the following NSF programs: Partnerships for Advanced
Computational Infrastructure, Distributed Terascale Facility (DTF) and
Terascale Extensions: Enhancements to the Extensible Terascale Facility.
\end{acknowledgements}

\clearpage
\bibliographystyle{apj}
\bibliography{myrefs}

\begin{thebibliography}{}

\bibitem[\protect\citeauthoryear{{Anninos} \& {Fragile}}{{Anninos} \&
  {Fragile}}{2003}]{ann03a}
{Anninos}, P.,  \& {Fragile}, P.~C. 2003, \apjs, 144, 243

\bibitem[\protect\citeauthoryear{{Anninos}, {Fragile}, \& {Murray}}{{Anninos}
  et~al.}{2003}]{ann03b}
{Anninos}, P., {Fragile}, P.~C.,  \& {Murray}, S.~D. 2003, \apjs, 147, 177

\bibitem[\protect\citeauthoryear{{Ant\'on}, {Zanotti}, {Miralles}, {Mart\'i},
  {Ib\'a\~nez}, {Font} \& {Pons}}{{Ant\'on} et~al.}{2005}]{anton05}
{Ant\'on}, L., {Zanotti}, O.,  {Miralles}, J.~A., {Mart\'i}, J.M.,
{Ib\'a\~nez}, J.~M., {Font}, J.~A. \& {Pons}, J.~A. 2005, astro-ph/0506063

\bibitem[\protect\citeauthoryear{{Balbus} \& {Hawley}}{{Balbus} \&
  {Hawley}}{1991}]{bal91}
{Balbus}, S.~A.,  \& {Hawley}, J.~F. 1991, \apj, 376, 214

\bibitem[\protect\citeauthoryear{{Berger} \& {Colella}}{{Berger} \&
  {Colella}}{1989}]{ber89}
{Berger}, M.~J.,  \& {Colella}, P. 1989, J. Comp. Phys., 82, 64

\bibitem[\protect\citeauthoryear{{Berger} \& {Oliger}}{{Berger} \&
  {Oliger}}{1984}]{ber84}
{Berger}, M.~J.,  \& {Oliger}, J. 1984, J. Comp. Phys., 53, 484

\bibitem[\protect\citeauthoryear{{Blandford} \& {McKee}}{{Blandford} \&
    {McKee}}{1976}]{bm76}
{Blandford}, R.~D., \& {McKee}, C.~F. 1976, Phys.Fluids, 19, 1130

\bibitem[\protect\citeauthoryear{{De Villiers} \& {Hawley}}{{De Villiers} \&
  {Hawley}}{2003a}]{dev03a}
{De Villiers}, J.,  \& {Hawley}, J.~F. 2003a, \apj, 589, 458

\bibitem[\protect\citeauthoryear{{De Villiers} \& {Hawley}}{{De Villiers} \&
  {Hawley}}{2003b}]{dev03b}
{De Villiers}, J.,  \& {Hawley}, J.~F. 2003b, \apj, 592, 1060

\bibitem[\protect\citeauthoryear{{Duez}, {Liu}, {Shapiro} \& {Stephens}}
  {{Duez} et~al.}{2005}]{duez05}
{Duez}, M.~D, {Liu}, Y.~T,  {Shapiro}, S.~L.,
\& {Stephens}, B.~C. 2005, astro-ph/0503420

\bibitem[\protect\citeauthoryear{{Fragile} \& {Anninos}}{{Fragile} \&
  {Anninos}}{2003}]{fra03a}
{Fragile}, P.~C.,  \& {Anninos}, P. 2003, \prd, 67, 103010

\bibitem[\protect\citeauthoryear{{Fragile} \& {Anninos}}{{Fragile} \&
  {Anninos}}{2005}]{fra05b}
{Fragile}, P.~C.,  \& {Anninos}, P. 2005, \apj, in press (astro-ph/0403356)

\bibitem[\protect\citeauthoryear{{Fragile} et~al.}{{Fragile}
  et~al.}{2005}]{fra05a}
{Fragile}, P.~C., {Anninos}, P., {Gustafson}, K.,  \& {Murray}, S.~D. 2005,
  \apj

\bibitem[\protect\citeauthoryear{{Gammie}, {McKinney}, \& {T{\' o}th}}{{Gammie}
  et~al.}{2003}]{gam03a}
{Gammie}, C.~F., {McKinney}, J.~C.,  \& {T{\' o}th}, G. 2003, \apj, 589, 444

\bibitem[\protect\citeauthoryear{{Hawley}, {Smarr}, \& {Wilson}}{{Hawley}
  et~al.}{1984}]{haw84b}
{Hawley}, J.~F., {Smarr}, L.~L.,  \& {Wilson}, J.~R. 1984, \apjs, 55, 211

\bibitem[\protect\citeauthoryear{{Hawley}, {Wilson}, \& {Smarr}}{{Hawley}
  et~al.}{1984}]{haw84a}
{Hawley}, J.~F., {Wilson}, J.~R.,  \& {Smarr}, L.~L. 1984, \apj, 277, 296

\bibitem[\protect\citeauthoryear{{Khokhlov}}{{Khokhlov}}{1998}]{kho98}
{Khokhlov}, A.~M. 1998, J. Comp. Phys., 143, 519

\bibitem[\protect\citeauthoryear{{Koide}, {Shibata}, \& {Kudoh}}{{Koide}
  et~al.}{1999}]{koi99}
{Koide}, S., {Shibata}, K.,  \& {Kudoh}, T. 1999, \apj, 522, 727

\bibitem[\protect\citeauthoryear{{Komissarov}}{{Komissarov}}{1999}]{kom99}
{Komissarov}, S.~S. 1999, \mnras, 303, 343

\bibitem[\protect\citeauthoryear{{Kurganov} \& {Tadmor}}{{Kurganov} \&
  {Tadmor}}{2000}]{kur00}
{Kurganov}, A.,  \& {Tadmor}, E. 2000, J. Comp. Phys., 160, 241

\bibitem[Landau \& Lifshitz(1959)]{ll59} Landau, L.~D., \& 
Lifshitz, E.~M.\ 1959, Fluid Mechanics, Course of theoretical physics, Oxford: Pergamon 
Press, 1959

\bibitem[\protect\citeauthoryear{{LeVeque}}{{LeVeque}}{1992}]{lev92}
{LeVeque}, R.~J. 1992, ``Numerical Methods for Conservation Laws'', 
(Birkhauser Verlag, Basel)

\bibitem[\protect\citeauthoryear{{Lucas-Serrano}, {Font},
  {Ib\'a\~nez} \& {Mart\'i}}{{Lucas-Serrano} et~al.}{2004}]{ser04}
{Lucas-Serrano}, A.,  {Font}, J.~A.,
{Ib\'a\~nez}, J.~M. \& {Marti\'i}, J.~M. 2004, A \& A, 428, 703 

\bibitem[\protect\citeauthoryear{{Michel}}{{Michel}}{1972}]{mic72}
{Michel}, F.~C. 1972, \apss, 15, 153

\bibitem[\protect\citeauthoryear{{Shu} \& {Osher}}{{Shu} \&
  {Osher}}{1988}]{shu88}
{Shu}, C.~W.,  \& {Osher}, S. 1988, J. Comp. Phys., 77, 439

\bibitem[\protect\citeauthoryear{{Stone} et~al.}{{Stone} et~al.}{1992}]{sto92}
{Stone}, J.~M., {Hawley}, J.~F., {Evans}, C.~R.,  \& {Norman}, M.~L. 1992,
  \apj, 388, 415

\bibitem[\protect\citeauthoryear{{Tscharnuter} \& {Winkler}}{{Tscharnuter} \&
  {Winkler}}{1979}]{tsc79}
{Tscharnuter}, W.~M.,  \& {Winkler}, K.~H. 1979, Comput. Phys. Comm., 18, 171

\bibitem[\protect\citeauthoryear{{von~Neumann} \& {Richtmyer}}{{von~Neumann} \&
  {Richtmyer}}{1950}]{von50}
{von~Neumann}, J.,  \& {Richtmyer}, R.~D. 1950, J. Appl. Phys., 21, 232

\bibitem[\protect\citeauthoryear{{White}}{{White}}{1973}]{whi73}
{White}, J.~W. 1973, J. Comp. Phys., 11, 573

\bibitem[\protect\citeauthoryear{{Wilson}}{{Wilson}}{1972}]{wil72}
{Wilson}, J.~R. 1972, \apj, 173, 431

\bibitem[\protect\citeauthoryear{{Wilson}}{{Wilson}}{1979}]{wil79}
{Wilson}, J.~R. 1979, in Sources of Gravitational Radiation, 423

\end{thebibliography}


\begin{table}[b]
\vskip30pt
\begin{tabular}{ccccc} \hline\hline
Grid  & Method  & $\Vert E(\rho) \Vert_1$  & $\Vert E(P) \Vert_1$ & $\Vert E(V) \Vert_1$  \\
  \hline
400      & AV   & $1.24\times 10^{-1}$  & $2.78\times 10^0$  & $1.38\times 10^{-2}$  \\
         & eAV  & $1.82\times 10^{-1}$  & $4.23\times 10^0$  & $1.99\times 10^{-2}$  \\
         & NOCD & $1.69\times 10^{-1}$  & $3.98\times 10^0$  & $2.00\times 10^{-2}$  \\
  \hline
800      & AV   & $8.09\times 10^{-2}$  & $1.61\times 10^0$  & $7.78\times 10^{-3}$  \\
         & eAV  & $9.00\times 10^{-2}$  & $2.06\times 10^0$  & $1.04\times 10^{-2}$  \\
         & NOCD & $1.04\times 10^{-1}$  & $2.00\times 10^0$  & $1.08\times 10^{-2}$  \\
  \hline
1600     & AV   & $5.03\times 10^{-2}$  & $1.01\times 10^0$  & $4.41\times 10^{-3}$  \\
         & eAV  & $5.18\times 10^{-2}$  & $1.01\times 10^0$  & $5.24\times 10^{-3}$  \\
         & NOCD & $6.59\times 10^{-2}$  & $1.02\times 10^0$  & $5.80\times 10^{-3}$  \\
  \hline
3200     & AV   & $3.83\times 10^{-2}$  & $7.12\times 10^{-1}$  & $2.35\times 10^{-3}$  \\
         & eAV  & $2.46\times 10^{-2}$  & $4.85\times 10^{-1}$  & $2.29\times 10^{-3}$  \\
         & NOCD & $3.14\times 10^{-2}$  & $5.17\times 10^{-1}$  & $2.71\times 10^{-3}$  \\
\hline\hline
\end{tabular}
\caption{$L$-1 Norm errors in density, pressure, and velocity for
the hydrodynamic shock-tube test at time $t=0.36$.
Convergence is approximately linear
as expected for problems with shock discontinuities.
}
\label{tab:errors1}
\end{table}

\begin{table}[b]
\vskip30pt
\begin{tabular}{cccccccccc} \hline\hline
  $\rho_1$     & $W_1$   & $W_r$  & $\rho_2$  &  $W_d$  & $e_d$   &$\rho_3$  &
  $W_f$     & $\rho_4$  & $W_4$ \\
  1.0       & 1.1436    & 5.412     & 14.0     &  3.0    & 56.0    & 14.0     &
  1.741     & 1.0       & 28.86 \\
  2.0       & 1.1436    & 4.482     & 24.28   &  2.572  & 79.06 & 16.20
  & 1.500  & 1.0       &  28.86 \\
  1.0       & 1.2558    & 9.973     & 26.5     &  5.0    & 238.5   & 26.5     &
  2.565   & 1.0       & 98.74 \\
\hline
\end{tabular}
\caption{Analytical results for boosted collisions shown in
Figs~\ref{fig:boostcol1}, \ref{fig:boostcol2} \& \ref{fig:boostcol3}
(see \S\ref{sec:boostcoll}).  Initial conditions are given by
$\rho_1$, $W_1$, $\rho_4$, $W_4$.  $W_d$ is solved for by Equations
(\ref{eqn:P2eP3}) \& (\ref{eqn:boostxforms}) in the center-of-momentum
frame, and then boosted into the lab frame.  Shocked proper densities
$\rho_2$ and $\rho_3$ are given by eqns.~(\ref{eqn:rho_boostshock}), and
the proper energy density of the contact discontinuity is $e_d =
\epsilon_2 \rho_2 = \epsilon_3 \rho_3$ with
eqns.~(\ref{eqn:eps_boostshock}). The boosts of the reverse and
forward shock fronts, $W_r$ and $W_f$, are calculated from their
center-of-momentum velocities (eqn.~(\ref{eqn:shockfronts})). }
\label{tab:boostcol}
\end{table}


\end{document}